\documentclass[twocolumn]{aastex701}
\usepackage[mathscr]{euscript}
\usepackage{threeparttable}
\usepackage{booktabs}
\usepackage{bm}% bold math
\usepackage{textcomp}
\usepackage{graphicx}% Include figure files
\usepackage{capt-of}
\usepackage{overpic}
\usepackage{multirow}
\usepackage{amsmath,amssymb,amsthm}
\usepackage{cases}
\usepackage{xcolor}
\usepackage[normalem]{ulem}
\expandafter\gdef\csname currCollabLimit0\endcsname{10000}
\shorttitle{Interacting dark energy after DESI DR2}
\shortauthors{Wang et al.}

\begin{document}

\title{Reassessing Evidence for Dark-Sector Interactions with Dynamical Dark Energy and DESI DR2}

\correspondingauthor{Hongwei Yu and Puxun Wu}
\email{hwyu@hunnu.edu.cn, pxwu@hunnu.edu.cn}

\author{Jincheng Wang}
\affiliation{Department of Physics, Key Laboratory of Low Dimensional Quantum Structures and Quantum Control of Ministry of Education, and Hunan Research Center of the Basic Discipline for Quantum Effects and Quantum Technologies, Hunan Normal University, Changsha, Hunan 410081, China}
\email{J.C.Wang@hunnu.edu.cn}

\author{Hongwei Yu}
\affiliation{Department of Physics, Key Laboratory of Low Dimensional Quantum Structures and Quantum Control of Ministry of Education, and Hunan Research Center of the Basic Discipline for Quantum Effects and Quantum Technologies, Hunan Normal University, Changsha, Hunan 410081, China}
\email{hwyu@hunnu.edu.cn}

\author{Puxun Wu}
\affiliation{Department of Physics, Key Laboratory of Low Dimensional Quantum Structures and Quantum Control of Ministry of Education, and Hunan Research Center of the Basic Discipline for Quantum Effects and Quantum Technologies, Hunan Normal University, Changsha, Hunan 410081, China}
\email{pxwu@hunnu.edu.cn}

\begin{abstract}
Recent baryon acoustic oscillation measurements from DESI Data Release 2, when combined with CMB and supernova data,  strengthen the motivation for exploring departures from $\Lambda$CDM in the late-time expansion history. 
Because an evolving dark-energy equation of state and an interaction within the dark sector can produce partially degenerate effects on the background expansion,  their observational signatures should be assessed simultaneously.  
We first consider  an interacting $w_0w_a$CDM model with $Q=\beta H\rho_{\rm de}$, using Planck and ACT CMB data, DESI DR2, and DES-Dovekie supernovae.  Allowing the dark-energy equation of state to evolve substantially weakens the preference for a nonzero coupling, while the preference for dynamical dark energy persists.  The interaction provides essentially no additional improvement in the best fit, suggesting that part of the coupling preference found in more restricted interacting models may reflect a degeneracy with dark-energy dynamics. We then examine the same interaction for three  one-parameter dynamical dark-energy trajectories: thawing, mirage, and generalized emergent dark energy (GEDE). The role of the interaction depends strongly on the assumed trajectory.  The thawing and GEDE models favor sizable couplings of opposite signs, whereas the mirage trajectory already closely follows the dark-energy evolution preferred by the data and provides little support for an additional interaction. These models also predict markedly different  signatures in structure growth, ranging from enhanced matter clustering in the interacting thawing model to strong suppression in interacting GEDE.  These contrasting growth signatures, despite substantial degeneracies at the background level, highlight late-time large-scale-structure observations as a promising avenue for distinguishing dark-energy dynamics from dark-sector interactions.

\end{abstract}


\section{Introduction}
\label{sec:intro}

The cosmological constant $\Lambda$, characterized by an equation of state $w=-1$, provides the simplest explanation for the observed late-time acceleration of the Universe.  Together with cold dark matter (CDM), it forms the standard $\Lambda$CDM cosmological model, which successfully accounts for a wide range of observations, such as cosmic microwave background (CMB) anisotropies~\citep{planck2018cosmo}, baryon acoustic oscillation (BAO) measurements~\citep{Eisenstein2005BAO,Alam2021eBOSS}, and type Ia supernova (SNIa) distances~\citep{riess1998observational,perlmutter1999measurements,Brout2022PantheonPlus}. However, the latest BAO measurements  from the Dark Energy Spectroscopic Instrument (DESI) Data Release 2 (DR2) have strengthened the motivation for considering  a time-dependent dark energy component~\citep{DESIDR2,Wang2025CosmicAcceleration,Li2026DESIReview}.  %Precisely because of this role as a baseline model, the recent DESI BAO indications for a time-dependent dark energy component provide a useful target for testing extensions beyond $\Lambda$CDM~\citep{Adame2024desi}.

 In the framework of the $w_0w_a$CDM cosmological model, in which the dark energy equation of state is described by the Chevallier-Polarski-Linder (CPL) parametrization $w(a)=w_0+w_a(1-a)$~\citep{chevallier2001accelerating,linder2003exploring}, the DESI BAO+CMB combination favors an extension beyond $\Lambda$CDM at about $3.1\sigma$, with the preference changing to $2.8$--$4.2\sigma$ when SNIa data are included~\citep{DESIDR2}. Here $w_0$ is the present-day value of $w(a)$, $w_a$ characterizes the time evolution of $w(a)$, and $a$ is the cosmic scale factor. The preferred parameter region generally has $w_0>-1$ and $w_0+w_a<-1$, which corresponds to an equation of state that evolves across the phantom divide, $w=-1$~\citep{DESI2025ExtendedDE}.  Motivated by these results, several alternative parameterizations have also been confronted with DESI BAO measurements, including  the Barboza--Alcaniz (BA)~\citep{BarbozaAlcaniz2008}, the Jassal--Bagla--Padmanabhan (JBP)~\citep{Jassal2005JBP}, the exponential (EXP)~\citep{Dimakis2016EXP,Pan2020EXP}, and the logarithmic (LOG)~\citep{Efstathiou1999LOG} forms. % have also been used to fit the DESI BAO data.  %Recently the DES-Dovekie recalibration leads to a more conservative Bayesian interpretation, but still preserves the same broad direction in the $w_0$-$w_a$ plane~\citep{desdovekie2025}. Thus, the current DESI distance data motivate tests of late time dark energy dynamics beyond the constant $w=-1$.
 
 A dynamical dark-energy component can arise naturally from a scalar field. The simplest example is quintessence~\citep{RatraPeebles1988,CaldwellLinder2005,Vikman2005PhantomCrossing}, for which a canonical kinetic term implies $w\geq-1$. Scalar-field models with $w<-1$, commonly referred to as phantom dark energy~\citep{Caldwell2002Phantom}, generally require a negative kinetic term and are consequently susceptible to quantum instabilities~\citep{CarrollHoffmanTrodden2003,Cline2004Phantom}.

An alternative mechanism capable of producing an effective phantom-like expansion history is an interaction between dark energy and dark matter~\citep{wetterich1995cosmon,amendola2000coupled,Wang2016Review,Wang2024FurtherIDE}. 
Such an interaction is conventionally described by an energy-transfer term $Q$ entering the continuity equations of CDM and dark energy.  By modifying the evolution of their energy densities, the interaction changes the Hubble expansion rate $H$ and can generate an effective dark-energy equation of state that crosses $w=-1$ even when the intrinsic dark-energy equation of state does not
%Thus, the dark matter-dark energy interaction alters the evolution of their energy densities, thereby modifying the Hubble expansion rate $H$. As a result, interacting dark energy (IDE) can facilitate the crossing of the phantom divide line in the effective equation of state for dark energy, even when the intrinsic equation of state is fixed at a value larger than or equal to $-1$
~\citep{AvelinoDaSilva2012,vonMarttens2019DarkDegeneracy,Chakraborty2025CoupledDESI,Antusch2026PhantomGuide}. %The interactions between cosmological constant dark energy and CDM (I$\Lambda$CDM), as well as those involving dark energy with a constant equation of state and CDM (I$w$CDM), have been extensively studied during the DESI BAO era
 Interacting models involving either vacuum energy (I$\Lambda$CDM) or dark energy with a constant equation of state (I$w$CDM) have therefore received renewed attention following the DESI measurements~\citep{Giare2024DESIIDE,LiEtAl2024DESIIDE,Silva2025DESIDR2IDE,Pan2026DESIDR2IDE,li2025updatedide,Zhu2026FermiDESIIDE,Li2026RobustIDE}. %Additionally, the couplings between quintessence and CDM have also been explored
Coupled quintessence scenarios have likewise been extensively investigated~\citep{wetterich1995cosmon,amendola2000coupled,Wang2026SUGRASignChange,Wang2026CQEXP,GomezValent2026DESICDE,Wang2026PlanckDESI,Chakraborty2026Chameleon}. %The results of these studies indicate that these models are capable of effectively explaining the DESI BAO measurements. 

%Both  a time-dependent dark energy equation of state and an interaction between the dark sectors can impact the late-time expansion history inferred from DESI BAO observations. At the background level, these two effects are partially degenerate %, since cosmological distance measurements constrain the total expansion history without uniquely determining the internal decomposition of the dark sector
An important complication is that a time-dependent dark-energy equation of state and a dark-sector interaction can produce partially degenerate effects on the late-time expansion history~\citep{Kunz2009DarkDegeneracy,Aviles2011DarkDegeneracy,petri2025darkdegeneracy,Guedezounme2026PhantomCrossing}. Cosmological distance measurements constrain the total expansion history but do not uniquely determine how the dark sector is decomposed into interacting dark matter and dark energy. 
Consequently, an apparent preference for a nonzero interaction obtained within a restricted framework, such as I$\Lambda$CDM or I$w$CDM, may  partly absorb an underlying time dependence of the dark-energy equation of state.  A central question is therefore whether evidence for a dark-sector interaction persists when the interaction and dark-energy dynamics are varied simultaneously~\citep{Yang2026BeyondDDE}, and how the answer depends on the assumed form of the dark-energy evolution.

%Thus, the critical question is whether the preference for interactions within the dark sector persists when both the interaction and the time dependence of the dark energy equation of state are explored simultaneously~\citep{Yang2026BeyondDDE}. Additionally, it is necessary to investigate how this conclusion varies with different assumptions regarding the dynamical dark energy model.

%Some studies have considered an interaction, proportional to $H\rho_{\rm de}$ with $H$ Hubble parameter and $\rho_{\rm de}$ energy density of dark energy, between the dark sectors with the equation of state of dark energy taking the parametrized forms, which allow the crossing of the phantom divide line~\citep{Shah2025InteractingDDE}, including CPL~\citep{Giare2024DynamicalIDE,Shah2025DESIDR2}, JBP~\citep{Jassal2005JBP,Shah2025DESIDR2}, and modified emergent dark energy~\citep{Benaoum2020MEDE}.  In these interacting analyses, however, the conventional fluid perturbation equations were evolved separately in the purely phantom and purely non-phantom sectors, so the phantom-divide-crossing region preferred by DESI was not sampled.
Several recent studies have considered interactions of the form $Q\propto H\rho_{\rm de}$, where $H$ is the Hubble parameter and $\rho_{\rm de}$ is the dark-energy density, together with phenomenological parametrizations of the dark-energy equation of state, including CPL, JBP, and modified emergent dark energy. However, \citet{Artola2026InteractingCPL,Neumann2026InteractingCPL} use only CMB distance priors, which primarily constrain the background geometry without directly incorporating the perturbation-level signatures of the interaction.  Other analyses include the full CMB spectra and dark-sector perturbations~\citep{Giare2024DynamicalIDE,Shah2025DESIDR2,Shah2025InteractingDDE}, but restrict the dark-energy equation of state to either the phantom or nonphantom regime.  It is  important to  revisit interacting dynamical dark-energy models using the full CMB spectra while allowing $w$ to evolve continuously across $-1$.
 
%These restriction reflect the difficulties in the conventional fluid treatment, mainly through two problem: the dark energy rest-frame pressure and velocity description becomes singular at $w=-1$, while the interaction can independently excite an early-time large-scale instability~\citep{FangHuLewis2008PPF,Valiviita2008,HeWangAbdalla2009}. To address both problems, the PPF framework was extended to interacting dark energy~\citep{li2014ppf,LiZhangZhang2014RSD,Zhang2017PPF}. Its model-independent implementation in IDECAMB enables stable perturbation evolution for interacting fluid models across the phantom divide, allowing them to be constrained with the full CMB spectra~\citep{li2023idecamb}.

In this work, we combine CMB temperature, polarization, and lensing measurements with DESI DR2 BAO and DES-Dovekie SNIa data to constrain interacting dynamical dark-energy models that allow phantom-divide crossing.   We adopt an interaction of the form $Q=\beta H\rho_{\rm de}$ and treat dark-energy perturbations using the generalized parameterized post-Friedmann framework, which permits stable evolution through $w=-1$.  We first consider the interacting CPL model, $Iw_0w_a$CDM. We find that allowing $w$ to evolve substantially weakens the preference for a nonzero coupling, whereas the preference for dynamical dark energy persists, in contrast to  the trend reported in previous analyses~\citep{Artola2026InteractingCPL,Neumann2026InteractingCPL,Shah2025DESIDR2}. Moreover, introducing the interaction provides essentially no additional improvement in the best-fit likelihood.

We then investigate whether this conclusion depends on the assumed dark-energy trajectory by considering three one-parameter models motivated by the DESI extended dark-energy analyses: thawing dark energy~\citep{CaldwellLinder2005,ScherrerSen2008,Chiba2009Thawing}, mirage dark energy~\citep{Linder2007Mirage}, and generalized emergent dark energy (GEDE)~\citep{LiShafieloo2020PEDE,YangEtAl2021GEDE}. We compare each model with its interacting counterpart and assess their relative statistical support using differences in the likelihood at the maximum a posteriori points and the Akaike information criterion (AIC). We further examine their predictions for the matter power spectrum, which provides a means of distinguishing models that remain partially degenerate at the level of the background expansion.

The remainder of this  paper is organized as follows. In Sec.~\ref{sec:model}, we introduce the interacting dark-sector framework and describe the observational data.  In Sec.~\ref{sec:results}, we present  the parameter constraints,  model comparisons, and implications for structure growth. We summarize our conclusions  in Sec.~\ref{sec:conclusion}.

\section{Models and Data}
\label{sec:model}

\subsection{Framework of interacting dark sectors}
\label{subsec:cf_model}

We consider a spatially flat universe in which baryons and radiation are separately conserved, while energy and momentum exchanges are allowed between CDM and dark energy~\citep{Valiviita2008,Clemson2012IDE,Wang2016Review,Wang2024FurtherIDE}. The total energy-momentum tensor satisfies the conservation equation
\begin{eqnarray}
 \nabla_\mu T^{\mu\nu}=0.
\label{eq:total_conservation}
\end{eqnarray} 
For the interacting dark sector, the individual conservation equations can be written as
\begin{eqnarray}
 \nabla_\mu T_{\rm c}^{\mu\nu}=-Q^{\nu},\quad
 \nabla_\mu T_{\rm de}^{\mu\nu}= Q^{\nu},
\label{Eq2}
\end{eqnarray}
where the subscripts `$\mathrm{c}$' and `$\mathrm{de}$' denote CDM and dark energy, respectively, and $Q^\nu$ characterizes the energy-momentum transfer between them. We take the transfer four-vector $Q^\nu$  to be parallel to the CDM four-velocity,
\begin{eqnarray}
  Q^\nu=Q u_{\rm c}^\nu,
\label{eq:q_four_vector}
\end{eqnarray}
such that the momentum transfer vanishes in the CDM rest frame. 

At the homogeneous and isotropic background level, Eq.~(\ref{Eq2})   reduces to
\begin{eqnarray}
 \dot{\rho}_c+3H\rho_c=-Q,\qquad
 \dot{\rho}_{\rm de}+3H(1+w)\rho_{\rm de}=Q,
\label{eq:dark_continuity}
\end{eqnarray}
where an overdot denotes a derivative with respect to cosmic time, $\rho_c$ and $\rho_{\rm de}$ are the CDM and dark-energy densities, respectively, and $w\equiv p_{\rm de}/\rho_{\rm de}$ is the intrinsic dark-energy equation of state. With this sign convention, $Q>0$ corresponds to energy transfer from CDM to dark energy, whereas $Q<0$ corresponds to transfer from dark energy to CDM. The noninteracting limit is recovered for $Q=0$.

In this paper, we focus on an interaction proportional to the  dark-energy density~\citep{Valiviita2008,Clemson2012IDE,li2014ppf,li2023idecamb}
\begin{eqnarray}
 Q=\beta H\rho_{\rm de},
\label{eq:q_beta}
\end{eqnarray}
where $\beta$ is a dimensionless coupling parameter. To illustrate the background-level degeneracy between the interaction and dark-energy dynamics, we define an effective dark-energy density
\begin{eqnarray}
 \rho_{\rm de}^{\rm eff}=\rho_{\rm de}+\rho_c-\rho_{c,0}a^{-3},
\label{eq:weff_de}
\end{eqnarray}
where $\rho_{c,0}$ is the present-day CDM density. Using Eqs.~\eqref{eq:dark_continuity} and \eqref{eq:weff_de}, one finds that $\rho_{\rm de}^{\rm eff}$ satisfies the standard noninteracting conservation equation
\begin{eqnarray}
    \dot{\rho}_{\rm de}^{\rm eff}+3H(1+w_{\rm de}^{\rm eff})\rho_{\rm de}^{\rm eff}=0
\end{eqnarray}
with $
 w_{\rm de}^{\rm eff}\equiv \frac{p_{\rm de}}{\rho_{\rm de}^{\rm eff}}$.
 Thus, at the background level, an interacting dark sector can be mapped onto a noninteracting model with an effective dark-energy equation of state. This illustrates why background distance measurements alone may have difficulty distinguishing dark-sector interactions from intrinsic dark-energy dynamics.

However, this equivalence does not generally extend to cosmological perturbations. The energy-momentum  transfer directly modifies the evolution of the CDM density contrast, $\delta_c\equiv\delta\rho_c/\rho_c$~\citep{Valiviita2008,Clemson2012IDE,li2014ppf,li2023idecamb}, which   satisfies
\begin{eqnarray}
 \delta_c'=-\frac{h'}{2}
 +\beta{\cal H}\frac{\rho_{\rm de}}{\rho_c}
 \left(\delta_c-\delta_{\rm de}\right),
\label{eq8}
\end{eqnarray}
where a prime denotes a derivative with respect to conformal time, $h$ is the trace of the scalar metric perturbation in the synchronous gauge,   $\delta_{\rm de}\equiv\delta\rho_{\rm de}/\rho_{\rm de}$, and ${\cal H}=aH$ . The term proportional to $\beta$ explicitly modifies the growth of CDM perturbations and vanishes in the noninteracting limit.
%describes the effect of the dark-sector interaction on the evolution of $\delta_c$. In the noninteracting limit, $\beta=0$, Eq.~(\ref{eq8}) reduces to the standard CDM perturbation equation. 

Since  Eq.~\eqref{eq8} explicitly depends  on $\delta_{\rm de}$,  dark-energy perturbations must also be evolved consistently.    In the conventional fluid description, the dark-energy density contrast obeys~\citep{Valiviita2008,Clemson2012IDE,li2023idecamb}
\begin{eqnarray}
 \delta_{\rm de}'={}&-(1+w)\left(kv_{\rm de}+\frac{h'}{2}\right)
 -3{\cal H}(c_{s,{\rm de}}^2-w)\delta_{\rm de}\nonumber\\
 &-3{\cal H}^2(c_{s,{\rm de}}^2-c_{a,{\rm de}}^2)
 \left[3(1+w)-\beta\right]\frac{v_{\rm de}}{k},
 \label{eq:de_density_perturbation}
 \end{eqnarray}
 where $c_{s,{\rm de}}^2$ and $c_{a,{\rm de}}^2$ are the rest-frame and adiabatic sound speeds squared, respectively, $k$ is the comoving wavenumber, and $v_{\rm de}$ is the dark-energy velocity potential, which satisfies the Euler equation
 \begin{eqnarray}
 v_{\rm de}'={}&-{\cal H}(1-3c_{s,{\rm de}}^2)v_{\rm de}
 +\frac{c_{s,{\rm de}}^2}{1+w}k\delta_{\rm de}\nonumber\\
 &-\frac{\beta{\cal H}(1+c_{s,{\rm de}}^2)}{1+w}v_{\rm de}.
 \label{eq:de_velocity_perturbation}
\end{eqnarray}
The factors of $1/(1+w)$ in Eq.~\eqref{eq:de_velocity_perturbation} render the conventional fluid description singular when $w$ crosses the phantom divide. Consequently, some previous analyses of interacting CPL- and JBP-type models treated the purely phantom and nonphantom regions separately~\citep{Giare2024DynamicalIDE,Shah2025InteractingDDE,Shah2025DESIDR2}.
%Equation~\eqref{eq:de_velocity_perturbation} contains terms proportional to $1/(1+w)$, and therefore becomes singular when $w$ crosses $-1$. To avoid this singularity, \citet{Giare2024DynamicalIDE,Shah2025InteractingDDE,Shah2025DESIDR2} considered the purely phantom and purely non-phantom regimes separately in their analyses of interacting CPL and JBP models. 

This restriction is particularly relevant in light of the DESI results, which favor regions of the $w_0$--$w_a$ plane in which the dark-energy equation of state can cross $w=-1$. To evolve the perturbations consistently through this crossing, we employ the parameterized post-Friedmann (PPF) framework,
generalized to interacting dark-energy models~\citep{FangHuLewis2008PPF,li2014ppf,LiZhangZhang2014RSD,Zhang2017PPF,li2023idecamb}. Rather than evolving the conventional dark-energy fluid variables through the singular point, the PPF approach provides a stable prescription for the large-scale relation between dark energy and the other cosmological components and smoothly connects it to the appropriate small-scale behavior. This allows us to explore interacting dynamical dark-energy models across the full parameter region, including trajectories that cross the phantom divide, while consistently incorporating their perturbation-level effects into the CMB observables.

\subsection{Observational data}
\label{subsec:data}

We constrain the cosmological models with a combination of CMB, BAO, and SNIa observations. The data sets are summarized below.

\begin{itemize}
\item \textbf{CMB}: We use the combined Planck and Atacama Cosmology Telescope data set, hereafter denoted as P-ACT. Following the ACT DR6 likelihood construction, the primary CMB spectra consist of the Planck low-multipole temperature and polarization likelihoods, the Planck high-multipole Plik lite likelihood restricted to $\ell_{\rm max}^{TT}=1000$ and $\ell_{\rm max}^{TE}=\ell_{\rm max}^{EE}=600$~\citep{Planck2018Likelihood}, and the compressed ACT DR6 CMB bandpower likelihood~\citep{ACTDR6Spectra}. This combination uses the large-angular-scale information from Planck together with the high-resolution ACT measurements of the small-angular-scale damping tail. We also include the ACT+Planck baseline CMB lensing likelihood, which combines the ACT DR6 lensing reconstruction with Planck lensing band powers~\citep{Qu2024ACTDR6LensingPower,ACTDR6Lensing,Carron2022PlanckPR4Lensing}.

\item \textbf{DESI BAO}:  We use the BAO measurements from DESI DR2~\citep{DESIDR2LyA,DESIDR2}, based on the first three years of DESI observations. These measurements combine galaxy, quasar, and Ly$\alpha$ forest tracers and constrain the distance-redshift relation over a broad range of redshifts.
%Recent BAO measurements have been released in DESI DR2~\citep{DESIDR2LyA,DESIDR2}. These measurements are based on the first three years of DESI observations and combine galaxy, quasar, and Lyman-$\alpha$ forest tracers to measure the distance-redshift relation over a broad redshift range.

\item \textbf{DES-Dovekie SNIa}: We use the DES-Dovekie SNIa sample~\citep{desdovekie2025}, a recalibrated analysis of the Dark Energy Survey five-year supernova sample~\citep{DES2024SN5YR}. The DES-Dovekie compilation incorporates improved photometric cross-calibration, updated light-curve training, and low-redshift supernova data.
\end{itemize}

Unless otherwise stated, the parameter constraints and model comparisons presented below are obtained from the combined P-ACT+DESI DR2+DES-Dovekie data set.

\section{Results and Discussion}
\label{sec:results}

We compute the CMB power spectra with a modified version of CAMB~\citep{LewisChallinorLasenby2000,li2023idecamb}. Cosmological parameter constraints are obtained with MCMC sampling using Cobaya~\citep{TorradoLewis2021}, and the resulting posterior distributions are analyzed with GetDist~\citep{Lewis2019GetDist}. 

To quantify the statistical performance of the models, we first consider the difference in $\chi^2$ evaluated at the maximum a posteriori (MAP) points,
\begin{eqnarray}
 \Delta\chi^2_{\rm MAP}\equiv-2\ln\left[ \mathcal{L}_{\rm MAP}({\rm model})/\mathcal{L}_{\rm MAP}(\Lambda{\rm CDM})\right],
\end{eqnarray}
where $\mathcal{L}_{\rm MAP}$ denotes the likelihood evaluated at the MAP point.  Negative values therefore indicate an improvement in the best fit relative to $\Lambda$CDM.  To account for differences in the number of free parameters, we also use  the Akaike information criterion (AIC)~\citep{Akaike1974}
\begin{eqnarray}
 \Delta{\rm AIC}=\Delta\chi^2_{\rm MAP}+2\Delta k,
\end{eqnarray}
where $\Delta k$ is the number of additional parameters relative to $\Lambda$CDM.  Thus, $\Delta{\rm AIC}<0$ favors the extended model, whereas $\Delta{\rm AIC}>0$ favors $\Lambda$CDM. As a commonly used rule of thumb, $|\Delta{\rm AIC}|\leq2$ provides little evidence for distinguishing the models, while $2<|\Delta{\rm AIC}|<6$, $6<|\Delta{\rm AIC}|<10$, and $|\Delta{\rm AIC}|>10$ are often interpreted as moderate, strong, and very strong support, respectively, for the model with the lower AIC~\citep{BurnhamAnderson2002}. These thresholds are used here only as qualitative guides for model comparison.

\subsection{\texorpdfstring{Interacting $w_0w_a$CDM}{Interacting w0waCDM}}
\label{subsec:results_iw0wa}

We begin with an  interacting extension of  $w_0w_a$CDM,  hereafter denoted as $Iw_0w_a$CDM,  with the dark-sector interaction  specified  by Eq.~\eqref{eq:q_beta}. This model allows both the intrinsic dark-energy equation of state and the interaction strength to vary and therefore provides a direct test of whether evidence for a dark-sector interaction persists once dynamical dark energy is simultaneously included.
For comparison, we also investigate the standard noninteracting $w_0w_a$CDM model. %The mean values along with $1\sigma$ uncertainties of model 
The marginalized parameter constraints  from the P-ACT+DESI DR2+DES-Dovekie  are summarized in Tab.~\ref{tab1}, and  the posterior distributions of $w_0$, $w_a$ and $\beta$ are shown in Fig.~\ref{fig1}.

\begin{center}
\centering
\includegraphics[width=0.42\columnwidth]{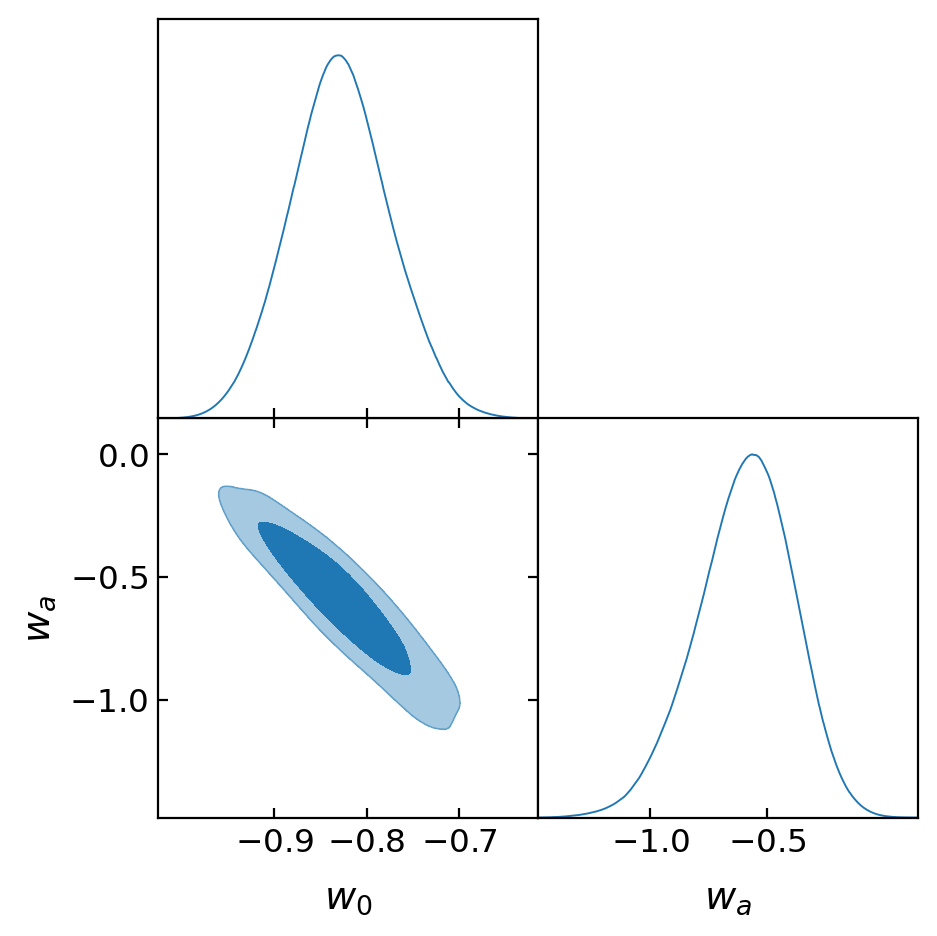}
\includegraphics[width=0.54\columnwidth]{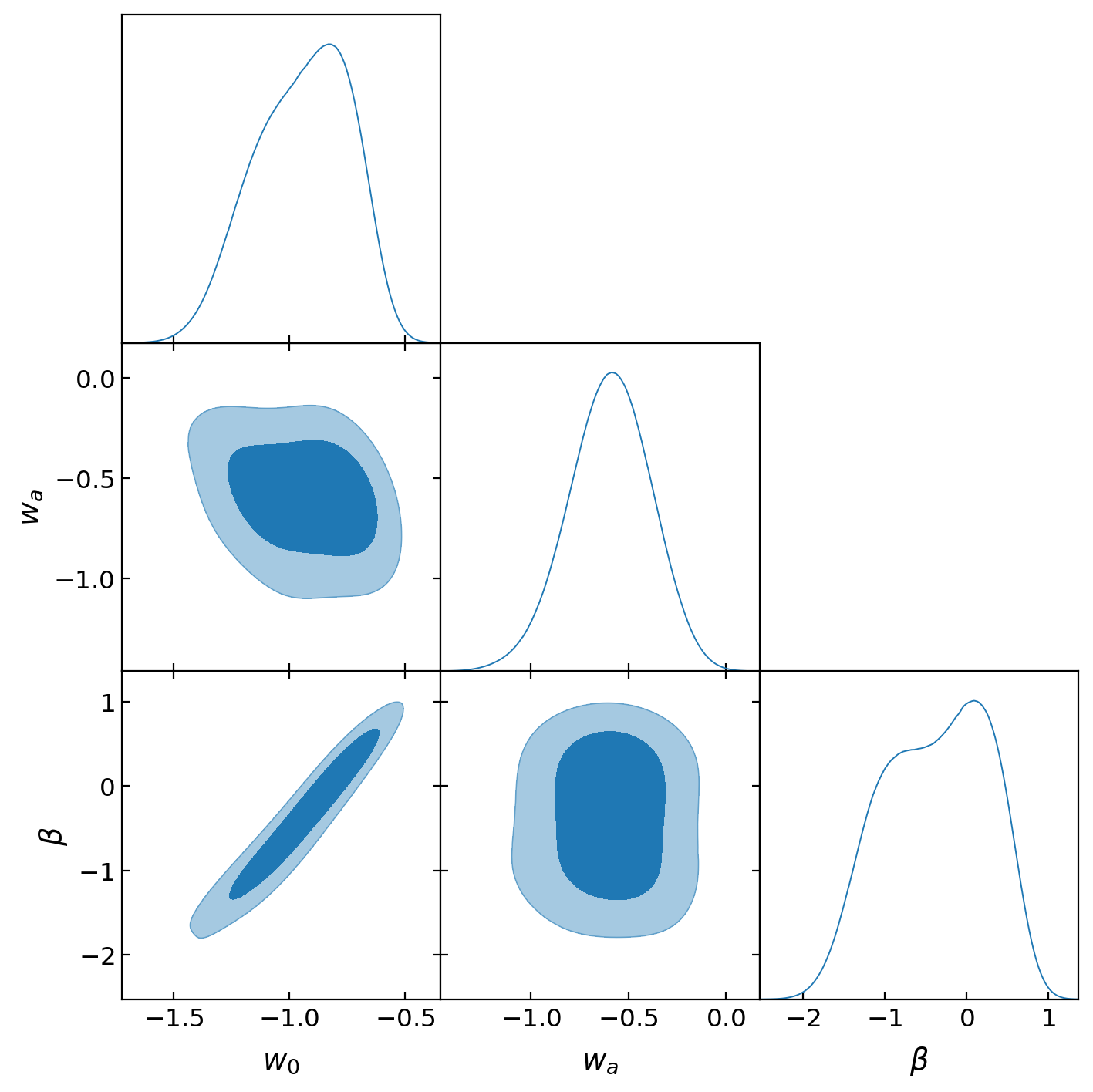}
\captionof{figure}{Posterior distributions of the model parameters for the   $w_0w_a$CDM  (left) and   $Iw_0w_a$CDM (right) models.}
\label{fig1}
\end{center}

\begin{center}
\centering
\captionof{table}{Marginalized parameter constraints (posterior means and $68\%$ credible intervals) from P-ACT+DESI DR2+DES-Dovekie. The last two rows give the differences in the MAP $\chi^2$ and AIC relative to $\Lambda$CDM.}
\label{tab1}
\small
\setlength{\tabcolsep}{3.5pt}
\begin{tabular}{lcc}
\hline\hline
Parameter & $w_0w_a$CDM & $Iw_0w_a$CDM \\
\hline
$w_0$ & $-0.829\pm0.054$ & $-0.93^{+0.24}_{-0.19}$ \\
$w_a$ & $-0.60^{+0.23}_{-0.18}$ & $-0.60\pm0.21$ \\
$\beta$ & $-$ & $-0.35^{+0.77}_{-0.62}$ \\
$H_0$ & $67.46\pm0.55$ & $67.40\pm0.54$ \\
$\Omega_{m0}$ & $0.3117\pm0.0053$ & $0.361^{+0.170}_{-0.098}$ \\
%$S_8$ & $0.8217\pm0.0071$ & $0.832^{+0.016}_{-0.160}$ \\
\hline\hline
$\Delta\chi^2_{\rm MAP}$ & $-10.72$ & $-10.75$ \\
$\Delta{\rm AIC}$ & $-6.72$ & $-4.75$ \\
\hline\hline
\end{tabular}
\end{center}

For the noninteracting $w_0w_a$CDM model, we obtain  $w_0=-0.829\pm0.054$ and $w_a=-0.60^{+0.23}_{-0.18}$, together with $\Omega_{m0}=0.3117\pm0.0053$ and $H_0=67.46\pm0.55\,{\rm km\,s^{-1}\,Mpc^{-1}}$. The inferred  $H_0$ remains substantially below the Cepheid-calibrated SNIa distance-ladder determination of $H_0=73.04\pm1.04\,{\rm km\,s^{-1}\,Mpc^{-1}}$ reported by the SH0ES collaboration~\citep{Riess2022SH0ES}; thus, allowing CPL-type dark-energy evolution does not alleviate the Hubble tension for the data combination considered here.

\begin{figure}[!tbp]
\centering
\includegraphics[width=0.8\columnwidth]{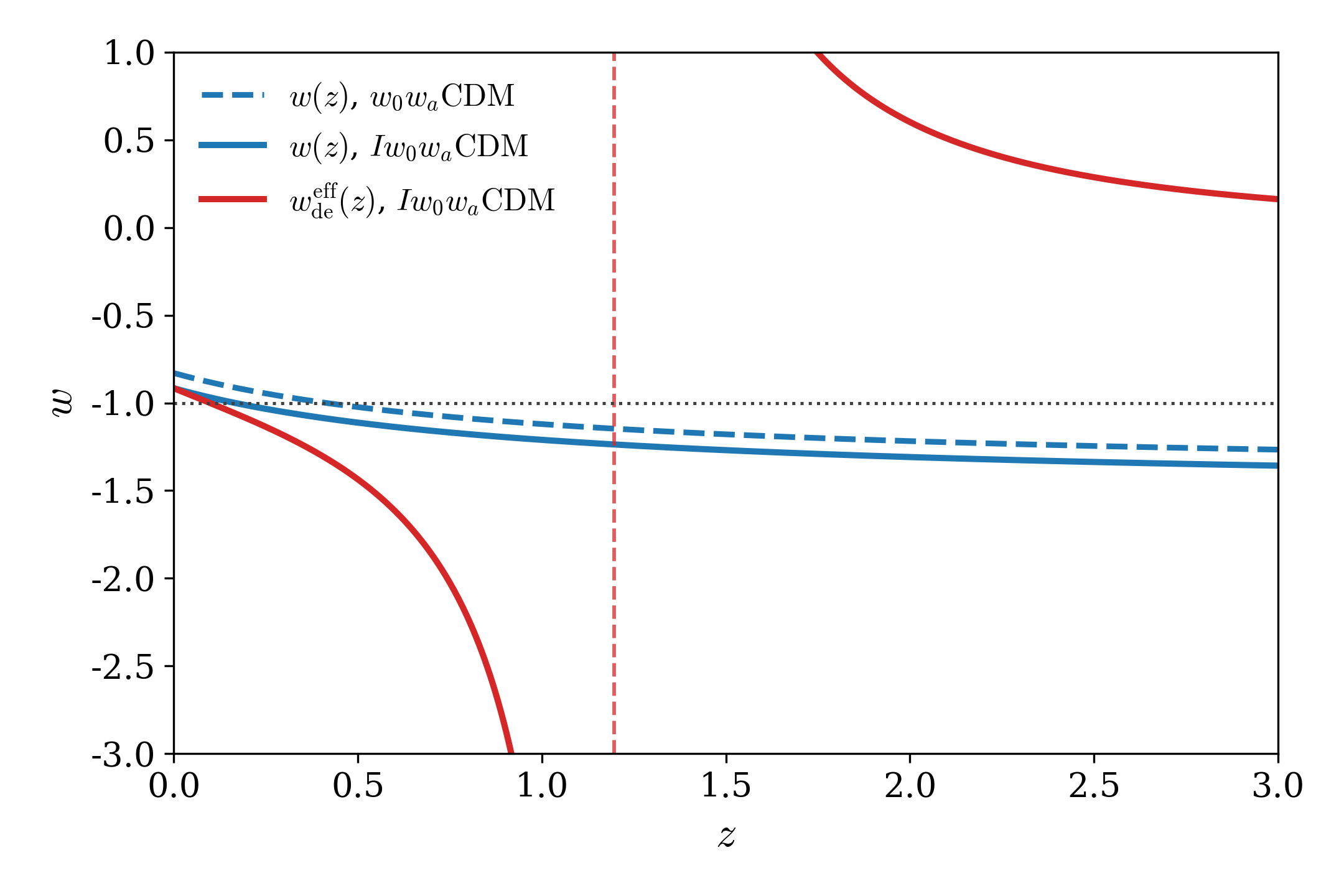}
\caption{Evolutions of the intrinsic  dark-energy equation of state $w(z)$ in $w_0w_a$CDM and $Iw_0w_a$CDM, together with the effective equation of state $w_{\rm de}^{\rm eff}(z)$ for the interacting model. The vertical dashed line marks  the redshift at which $\rho_{\rm de}^{\rm eff}=0$ and $w_{\rm de}^{\rm eff}$ develops a pole.}
\label{fig2}
\end{figure}

The preferred dark-energy parameters correspond to $w_0>-1$ and $w_0+w_a<-1$. As illustrated in Figure~\ref{fig2}, the resulting equation of state evolves from a phantom-like regime at earlier times to a quintessence-like regime at late times, crossing $w=-1$ during the evolution. This behavior is consistent with the qualitative trend reported by DESI~\citep{DESIDR2}. Relative to $\Lambda$CDM, the model improves the best-fit statistic by $\Delta\chi^2_{\rm MAP}=-10.72$. After accounting for its two additional parameters, $\Delta{\rm AIC}=-6.72$, corresponding to strong support for $w_0w_a$CDM over $\Lambda$CDM according to the conventional AIC criterion.

%corresponding to a $4.8\sigma$ discrepancy. The dark-energy parameters are constrained to $w_0=-0.829\pm0.054$ and $w_a=-0.60^{+0.23}_{-0.18}$. The constraint on $w_a$ excludes zero at more than $2\sigma$,  favoring  an evolving dark-energy equation of state that crosses the phantom divide from   $w<-1$ to $w>-1$ during cosmic expansion, as illustrated in Fig.~\ref{fig2}. This behavior is consistent with the DESI results~\citep{DESIDR2}. Relative to $\Lambda$CDM, the $w_0w_a$CDM model yields $\Delta\chi^2_{\rm MAP}=-10.72$, indicating a substantial improvement in the fit, while  $\Delta{\rm AIC}=-6.72$ indicates a strong preference for dynamical dark energy. 

We next allow for a simultaneous dark-sector interaction. The coupling
%After extending the $w_0w_a$CDM model to include a dark sector interaction, we find from Tab.~\ref{tab1} that the coupling parameter $\beta$ 
is constrained to $\beta=-0.35^{+0.77}_{-0.62}$, and is therefore consistent with zero at $1\sigma$. The posterior provides no compelling evidence for a nonzero interaction.  This result  contrasts with the coupling preference 
%for a nonzero coupling found in the interacting $\Lambda$CDM analyses
reported in some more restricted interacting models ~\citep{Giare2024DESIIDE,LiEtAl2024DESIIDE,Pan2026DESIDR2IDE,li2025updatedide}, suggesting that part of the apparent interaction signal in such models may be absorbed by allowing additional freedom in the intrinsic dark-energy evolution.

%the preference for an interaction is weakened once dynamical dark energy is allowed.
An important consequence of introducing the interaction is the substantial broadening of the matter-density constraint. We obtain $\Omega_{m0}=0.361^{+0.170}_{-0.098}$, compared with $\Omega_{m0}=0.3117\pm0.0053$  in the noninteracting $w_0w_a$CDM model.  Although the posterior mean shifts upward, the two determinations remain statistically consistent because of the much larger uncertainty in the interacting model.  This broadening reflects a strong degeneracy in the decomposition of the dark sector: changes in the present-day matter abundance can be partially compensated by changes in the coupling and in the dark-energy equation of state while preserving a similar expansion history.

%However, these two results remain consistent  within $1\sigma$ because of the substantially increased uncertainty in the interacting case. 
By contrast,   $H_0$ and $w_a$ remain remarkably stable: $H_0=67.40\pm0.54\,{\rm km\,s^{-1}\,Mpc^{-1}}$ and $w_a=-0.60\pm0.21$.  The interaction therefore neither appreciably changes the inferred Hubble constant nor removes the preference for an evolving dark-energy equation of state. The central value of $w_0$ also remains statistically consistent with the noninteracting result, although its uncertainty increases considerably. As shown in Figure~\ref{fig1}, this loss of constraining power is associated in part with a pronounced correlation between $w_0$ and $\beta$.

%, indicating that it does not alleviate the Hubble tension.  Moreover, $w_a$ remains inconsistent  with zero at more than $2\sigma$, showing   that the preference for an evolving $w(z)$ persists  even in the presence of the dark-sector interaction, as illustrated  in Fig.~\ref{fig2}.   The central value of $w_0$ also  remains consistent with the noninteracting result, although its uncertainty increases substantially. As shown in Fig.~\ref{fig1}, $\beta$ and $w_0$ exhibit a noticeable positive correlation.  

Figure~\ref{fig2} compares the intrinsic equation of state $w(z)$ with the effective equation of state $w_{\rm de}^{\rm eff}(z)$.  For the posterior mean parameters of the interacting model, $w_{\rm de}^{\rm eff}$ develops a pole at $z\simeq1.20$, where the reconstructed effective density  $\rho_{\rm de}^{\rm eff}$ crosses zero. This behavior follows directly  from the modified CDM dilution history.  For   the central value  $\beta<0$, energy is transferred  from dark energy to CDM, causing the CDM density  to dilute more slowly than $a^{-3}$ as the Universe expands. Equivalently,  after normalizing the interacting and reference noninteracting histories  to the same present-day CDM density $\rho_{c,0}$,  the interacting CDM density at earlier times satisfies  $\rho_c<\rho_{c,0}a^{-3}$.  The second and third terms in Eq.~\eqref{eq:weff_de} therefore give a negative contribution to $\rho_{\rm de}^{\rm eff}$. At $z\simeq1.20$, this contribution cancels the intrinsic dark-energy density, $\rho_{\rm de}=\rho_{c,0}a^{-3}-\rho_c$, so that $\rho_{\rm de}^{\rm eff}=0$ and $w_{\rm de}^{\rm eff}=p_{\rm de}/\rho_{\rm de}^{\rm eff}$ diverges.

The pole should not be interpreted as a physical singularity. Both $\rho_{\rm de}$ and $\rho_c$, as well as the underlying expansion history, remain regular. Rather, the divergence signals a breakdown of the effective single-fluid equation-of-state description when its reconstructed energy density passes through zero. The associated phantom-like excursion at lower redshift and positive values of $w_{\rm de}^{\rm eff}$ on the high-redshift side of the pole are therefore properties of the effective noninteracting representation rather than pathologies of the underlying interacting model.

%the CDM density in the past  is smaller in the interacting  case, such that $\rho_c<\rho_{c,0}a^{-3}$.  Consequently, the term $\rho_c-\rho_{c,0}a^{-3}$ in Eq.~\eqref{eq:weff_de} becomes negative. At $z\simeq1.20$, its magnitude equals the dark-energy density, i.e., $\rho_{\rm de}=\rho_{c,0}a^{-3}-\rho_c$, so that $\rho_{\rm de}^{\rm eff}=0$ and the reconstructed effective equation of state develops a pole. At lower redshifts, $w_{\rm de}^{\rm eff}$ exhibits a more pronounced excursion into the phantom regime, whereas it becomes positive on the high-redshift side of the pole. Importantly, both the pole and these unusual features arise from the effective noninteracting representation of the interacting dark sector and do not correspond to a physical singularity of the underlying IDE model.

Most importantly, introducing the coupling produces essentially no improvement in the maximum likelihood. 
We find $\Delta\chi^2_{\rm MAP}=-10.75$ for $Iw_0w_a$CDM, compared with $-10.72$ for $w_0w_a$CDM.  The additional interaction parameter therefore improves $\chi^2_{\rm MAP}$ by only $0.03$. Once the extra degree of freedom is penalized, $\Delta{\rm AIC}$ changes from $-6.72$ to $-4.75$. Thus, for the data considered here, the improvement over $\Lambda$CDM is driven primarily by the freedom to evolve the dark-energy equation of state rather than by the dark-sector interaction. In this sense, allowing dynamical dark energy substantially reduces the statistical motivation for an additional coupling.

%After accounting for the additional parameter, $\Delta{\rm AIC}$ increases from $-6.72$ to $-4.75$, reducing the statistical preference for the interacting extension. Thus,  the data continue to favor dynamical dark energy but provide little statistical support for an additional dark-sector interaction.

The preceding results demonstrate that dark-energy dynamics and dark-sector interactions can be strongly degenerate in background observables.
%We have shown that dynamical dark energy and dark-sector interaction can produce partially degenerate effects under the observational constraints considered here.  %indicating that current observations lack the sensitivity needed to robustly discriminate between these two models.
% However, they can lead to distinct signatures  in structure growth
 Their perturbation-level effects, however, need not be degenerate~\citep{PourtsidouTram2016,DiValentino2020IDE,Zhai2023CMBIDE}.  In a non-interacting model,  dark energy affects matter clustering primarily  through the background expansion and, depending on its clustering properties, through its own perturbations.   An interaction additionally modifies the background evolution of CDM and directly enters its perturbation equation. Measurements of structure growth can therefore provide information complementary to that contained in distance observables.
 %whereas a dark sector interaction can directly  modify both the background and the perturbation evolution of CDM. The matter power spectrum  therefore provides a useful probe for breaking this degeneracy and distinguishing their effects on structure formation. 
 The matter power spectrum characterizes the variance of the total matter density contrast in Fourier space,
\begin{eqnarray}
    P_{m}(k)\propto
    \left\langle |\delta_{m,{\bf k}}|^2\right\rangle,
\label{eq:pm_delta_m}
\end{eqnarray}
where
\begin{eqnarray}
    \delta_{m,{\bf k}}=\frac{\rho_b\delta_{b,{\bf k}}+\rho_c\delta_{c,{\bf k}}+\rho_\nu\delta_{\nu,{\bf k}}}{\rho_b+\rho_c+\rho_\nu}.
\end{eqnarray}
Here ${\bf k}$ is the comoving wave vector, $k=|{\bf k}|$, and $\delta_{b,{\bf k}}$, $\delta_{c,{\bf k}}$, and $\delta_{\nu,{\bf k}}$ denote the Fourier modes of the baryon, CDM, and massive-neutrino density contrasts, respectively. This expression makes clear that CDM enters $P_{m}(k)$ through both its weight, $\rho_c/(\rho_b+\rho_c+\rho_\nu)$, and its density contrast mode $\delta_{c,{\bf k}}$.

\begin{figure}[!tbp]
\centering
\includegraphics[width=0.8\columnwidth]{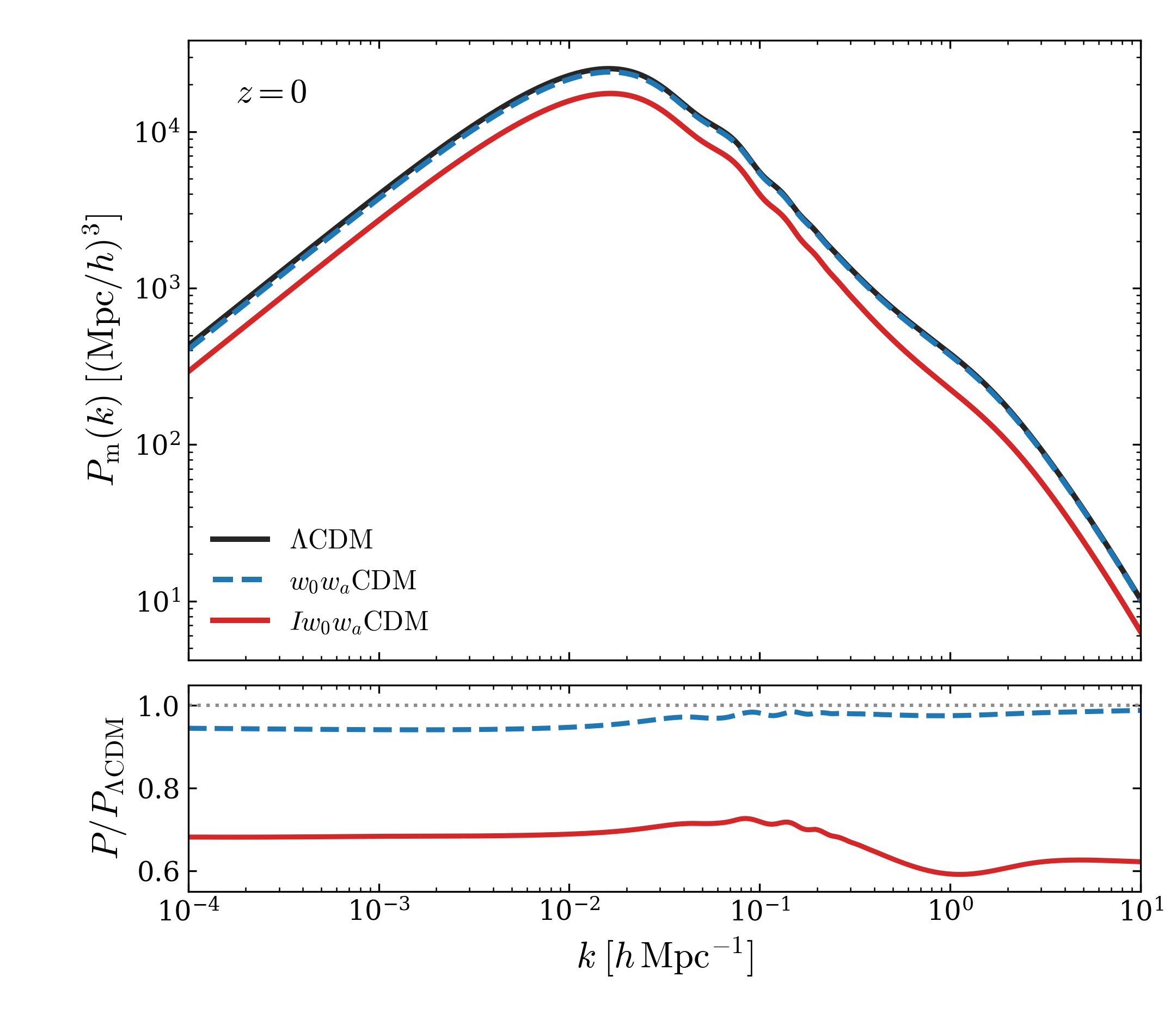}
\caption{Present-day matter power spectra for $w_0w_a$CDM and $Iw_0w_a$CDM. The lower panel shows the ratio to the corresponding $\Lambda$CDM spectrum.}
\label{fig3}
\end{figure}

Fig.~\ref{fig3} shows the present-day matter power spectra and their ratios to $\Lambda$CDM~\citep{Mead2021HMCode2020}. The noninteracting $w_0w_a$CDM spectrum is close to that of $\Lambda$CDM, indicating that the preferred noninteracting dynamical-dark-energy history  produces only a modest change in present-day matter clustering. By contrast, the  $Iw_0w_a$CDM spectrum is substantially suppressed over the plotted range of scales. Thus, models that provide very similar fits to the background expansion can nevertheless predict appreciably different growth histories.

% demonstrating that the dark-sector interaction has a significant impact on structure growth.

This behavior can be understood from the evolution of the CDM density perturbation. Eq.~\eqref{eq8} can  be expressed in Fourier space as
\begin{eqnarray}
 \delta_{c,{\bf k}}'=(\delta_{c,{\bf k}}')_{\beta=0}
 +\beta{\cal H}\frac{\rho_{\rm de}}{\rho_c}
 \left(\delta_{c,{\bf k}}-\delta_{{\rm de},{\bf k}}\right),
\label{eq:delta_c_ide}
\end{eqnarray}
where $(\delta_{c,{\bf k}}')_{\beta=0}$ denotes the standard noninteracting contribution to the CDM perturbation evolution.  On subhorizon scales,   dark-energy perturbations  are typically  negligible compared with matter perturbations~\citep{FangHuLewis2008PPF,li2014ppf,planck2018cosmo}, so that the interaction term approximately reduces to $\beta{\cal H}\frac{\rho_{\rm de}}{\rho_c}
 \left(\delta_{c,{\bf k}}-\delta_{{\rm de},{\bf k}}\right)\simeq\beta{\cal H}\frac{\rho_{\rm de}}{\rho_c} \delta_{c,{\bf k}}$.  For the negative central value of $\beta$ in the $Iw_0w_a$CDM fit,  this contribution acts to reduce
 the growth of $\delta_{c,{\bf k}}$.  Together with the interaction-induced modification of the CDM background density, this leads to a suppression of $\delta_m$ and hence of $P_m(k)$.
 
 Because $\beta$ is only weakly constrained in $Iw_0w_a$CDM, however, the spectrum shown in Figure~\ref{fig3} should be interpreted with some care. It illustrates the growth history associated with the representative parameter set used to generate the figure rather than establishing a precisely determined prediction for the entire interacting posterior. A full assessment of the allowed range of structure-growth histories would require propagating the posterior distribution into observables such as $P_m(k)$, $f\sigma_8(z)$, or weak-lensing statistics.
 
  %Through Eq.~\eqref{eq:pm_delta_m},  the reduced CDM density contrast in turn suppresses  the total  matter perturbation $\delta_{m,{\bf k}}$ and hence the matter power spectrum.  The suppression of $P_{m}(k)$ in the interacting model can  therefore be traced directly to the negative coupling.

Although  the representative  interacting model produces a visible change in the present-day matter power spectrum, its impact on the CMB lensing spectrum  is comparatively modest for the data combination considered here.
%remains relatively small. Current CMB lensing data may therefore have limited sensitivity to this effect.  
Low-redshift large-scale-structure measurements that probe the matter distribution more directly may therefore provide substantially greater discriminatory power.   Galaxy clustering, redshift-space distortions, and weak gravitational lensing are particularly promising for determining whether the growth histories permitted by the background data remain observationally viable.

%such as galaxy clustering measurements from the DESI Legacy Imaging Survey, could provide additional discriminatory power between interacting and noninteracting dynamical dark-energy models~\citep{Dey2019Legacy,Hang2021LegacyClustering}.
 It is useful to compare these results with previous analyses of interacting CPL models. \citet{Artola2026InteractingCPL} considered the same background interaction,  $Q\propto H\rho_{\rm de}$, within the CPL framework using compressed Planck+ACT CMB information, DESI DR2, and Pantheon+. Despite the different SNIa sample, their noninteracting CPL constraints closely agree with ours,  providing a useful common baseline. When the interaction was included, however, their preferred value of $w_a$ shifted markedly toward zero, accompanied by a mild preference for energy transfer from dark energy to CDM, thereby weakening the inferred  evidence of the dark-energy evolution. 
 
 Our result differs in an important respect. When the CMB temperature and polarization spectra and CMB lensing information are included, the preferred $w_a$ remains nearly unchanged after introducing the interaction, while $\beta$ remains consistent with zero and the uncertainties in $w_0$ and $\Omega_{m0}$ increase substantially. We therefore find that the additional freedom associated with the interaction is expressed primarily through broadened parameter degeneracies rather than through a shift of the preferred dark-energy evolution toward $w_a=0$.
 
 %By contrast, in our analysis using the CMB temperature and polarization spectra together with CMB lensing, $w_a$ and $H_0$ remain nearly unchanged, while the coupling is consistent with zero and the constraints on $w_0$ and $\Omega_{m0}$ broaden. We therefore do not recover the coupling-induced weakening of the dynamical dark-energy preference  found in the compressed-CMB analysis.

A complementary comparison can be made with \citet{Shah2025DESIDR2}, who analyzed the phantom and non-phantom CPL regimes separately using Planck, DESI DR2, and Pantheon+. In the phantom branch, energy transfer from dark energy to CDM is accompanied by a higher matter density and substantially suppressed clustering, while the equation of state approached $-1$ from below at late times. The non-phantom branch  exhibits the opposite  direction of energy transfer and enhanced structure growth.  Our analysis instead permits the intrinsic equation of state to evolve continuously across $w=-1$. Within this enlarged parameter space, the data retain a preference for dynamical dark energy but do not require a nonzero interaction. Both analyses nevertheless find that introducing the interaction has little effect on the inferred value of $H_0$.

%By allowing the equation of state to cross the phantom divide, our analysis instead retains the preference for dynamical dark energy without finding either a strong preference for the interaction or a comparably large shift in the matter sector. Both analyses, however, find little impact on  $H_0$.

Taken together, these results indicate that the inferred evidence for a dark-sector interaction depends sensitively on the freedom allowed in the dark-energy sector and on the treatment of cosmological perturbations. The persistence of the dynamical-dark-energy preference in $Iw_0w_a$CDM motivates us to ask whether the conclusion changes when the dark-energy evolution is restricted to theoretically or phenomenologically motivated trajectories. We therefore turn next to three one-parameter models considered in DESI extended dark-energy analyses---thawing, mirage, and GEDE---and examine how their cosmological constraints and structure-growth predictions change when the same dark-sector interaction is introduced~\citep{Lodha2024PhysicsDE,DESI2025ExtendedDE}.

%Motivated by the persistent preference for dynamical dark energy in the interacting $w_0w_a$CDM analysis, we next consider three specific one-parameter trajectories explored in the DESI analyses---thawing, mirage, and GEDE---and investigate how their constraints are affected by the dark-sector interaction~\citep{Lodha2024PhysicsDE,DESI2025ExtendedDE}.

\subsection{Thawing dark energy}
\label{subsec:thawing_results}

We next consider the thawing class of dark-energy models. In canonical thawing scenarios, the scalar field is initially frozen by Hubble friction and begins to evolve only at relatively late times~\citep{CaldwellLinder2005,ScherrerSen2008,Chiba2009Thawing}. Following the calibrated trajectory adopted in the DESI extended dark energy analyses~\citep{Lodha2024PhysicsDE,DESI2025ExtendedDE}, we describe  this class in the $w_0$-$w_a$ plane by
\begin{eqnarray}
 w_a=-1.58(1+w_0).
\label{eq:thawing_model}
\end{eqnarray}
The dark-energy equation of state is therefore specified by the single  independent parameter $w_0$.  We emphasize that Eq.~\eqref{eq:thawing_model} is used here as a phenomenological one-parameter trajectory in the $w_0$--$w_a$ plane rather than as a complete microscopic scalar-field model. In particular,  depending on the value of $w_0$, its CPL continuation can extend into the phantom regime. The interacting extension introduces the additional coupling parameter $\beta$ through Eq.~\eqref{eq:q_beta}. The posterior distributions are shown in Fig.~\ref{fig4}, and the marginalized constraints are listed in Tab.~\ref{tab:thawing_constraints}.

\begin{center}
\centering
\includegraphics[width=0.42\columnwidth]{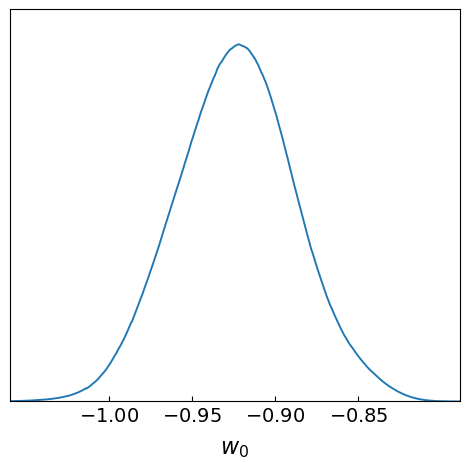}
\includegraphics[width=0.54\columnwidth]{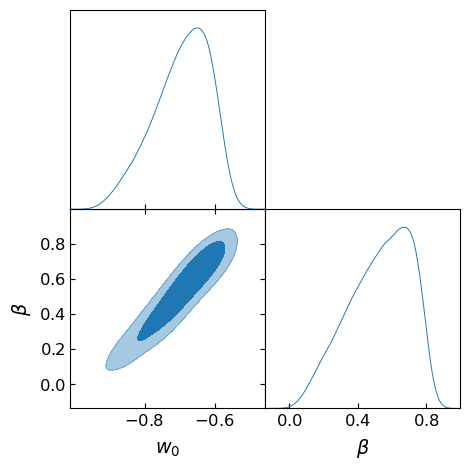}
\captionof{figure}{Posterior distributions of selected parameters for the thawing model and its interacting extension.}
\label{fig4}
\end{center}

\begin{center}
\centering
\captionof{table}{Constraints for the thawing model and its interacting extension.}
\label{tab:thawing_constraints}
\small
\setlength{\tabcolsep}{3.5pt}
\begin{tabular}{lcc}
\hline\hline
Parameter & Thawing & Interacting thawing \\
\hline
$w_0$ & $-0.923\pm0.036$ & $-0.692^{+0.100}_{-0.060}$ \\
$\beta$ & $-$ & $0.53^{+0.24}_{-0.13}$ \\
$H_0$ & $67.33\pm0.56$ & $67.43\pm0.56$ \\
$\Omega_{m0}$ & $0.3093\pm0.0055$ & $0.156^{+0.050}_{-0.077}$ \\
%$S_8$ & $0.8140\pm0.0074$ & $1.14^{+0.10}_{-0.27}$ \\
\hline\hline
$\Delta\chi^2_{\rm MAP}$ & $-3.93$ & $-10.00$ \\
$\Delta{\rm AIC}$ & $-1.93$ & $-6.00$ \\
\hline\hline
\end{tabular}
\end{center}

\begin{figure}[!htbp]
\centering
\includegraphics[width=0.8\columnwidth]{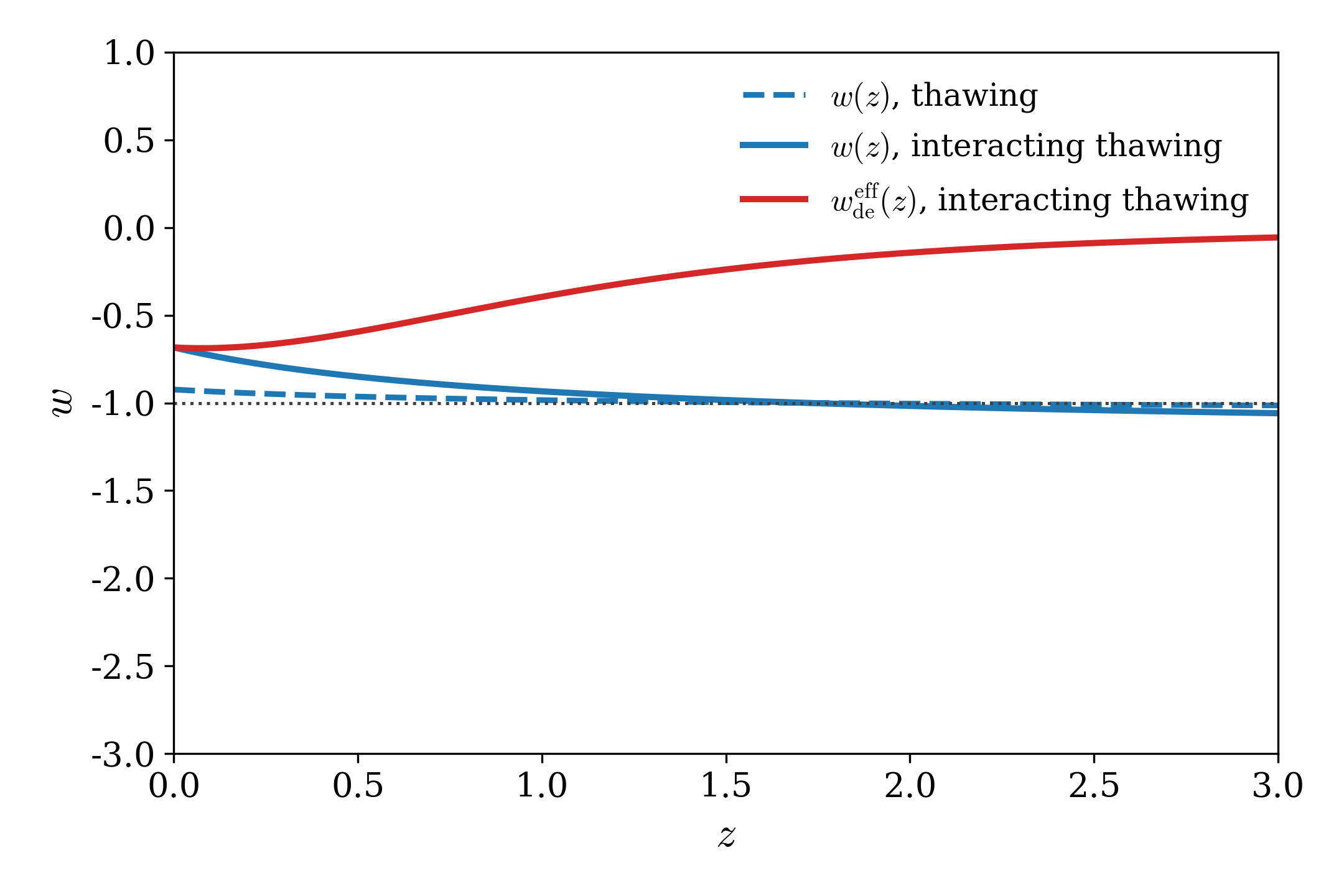}
\caption{Evolutions of the intrinsic dark-energy equation of state for the thawing model and its interacting extension.}
\label{fig5}
\end{figure}

For the noninteracting thawing model, we obtain $w_0=-0.923\pm0.036$, together with $\Omega_{m0}=0.3093\pm0.0055$ and $H_0=67.33\pm0.56\,{\rm km\,s^{-1}\,Mpc^{-1}}$. Through Eq.~\eqref{eq:thawing_model}, the preferred value of $w_0$ corresponds to $w_a<0$. As shown in Figure~\ref{fig5}, the resulting equation of state therefore evolves from a phantom-like regime at earlier times toward $w_0>-1$ at the present epoch, qualitatively resembling the evolution preferred in the full $w_0w_a$CDM model but with its time dependence restricted to the one-dimensional thawing trajectory.

Relative to $\Lambda$CDM, the noninteracting thawing model gives $\Delta\chi^2_{\rm MAP}=-3.93$ and $\Delta{\rm AIC}=-1.93$. The improvement in the best fit is therefore insufficient to provide meaningful evidence for the thawing trajectory over $\Lambda$CDM once the additional parameter is taken into account.

%For the noninteracting thawing model, the constraints on $\Omega_{m0}$ and $H_0$ are consistent with those reported in Tab.~\ref{tab1}. The model parameter $w_0$ is constrained to $w_0=-0.923\pm0.036$, with a smaller uncertainty than that obtained in the $w_0w_a$CDM model. According to Eq.~\eqref{eq:thawing_model}, this corresponds to  $w_a<0$, resembling the behavior observed in the $w_0w_a$CDM model. As shown in Fig.~\ref{fig5}, the  resulting  $w(z)$ therefore retains the tendency toward an evolving dark-energy equation of state with phantom-divide crossing,  but  with a more restricted time dependence. Relative to $\Lambda$CDM, the thawing model yields $\Delta\chi^2_{\rm MAP}=-3.93$ and $\Delta{\rm AIC}=-1.93$. These results indicate that there is no significant preference for the thawing model over $\Lambda$CDM. 

The situation changes substantially when the dark-sector interaction is introduced. We obtain that $\beta=0.53^{+0.24}_{-0.13}$, with the posterior favoring a positive coupling.
%indicating a preference for a positive coupling   at more than $2\sigma$. 
Under our sign convention, $\beta>0$   corresponds to  energy transfer  from CDM to dark energy. At the same time, the preferred present-day equation of state   shifts from $w_0=-0.923\pm0.036$ to $w_0=-0.692^{+0.100}_{-0.060}$.  Because $w_a=-1.58(1+w_0)$, this shift implies a substantially more negative $w_a$ and hence a stronger evolution of the intrinsic equation of state than in the noninteracting thawing model.

Figure~\ref{fig5} illustrates the corresponding evolution. Although the intrinsic $w(z)$ evolves strongly and crosses the phantom divide, the reconstructed effective equation of state $w_{\rm de}^{\rm eff}$ remains above $-1$ over the redshift range shown. The interaction therefore changes not only the preferred parameters but also the effective interpretation of the dark-energy evolution. This provides a concrete example of the background-level degeneracy discussed in Sec.~\ref{subsec:cf_model}: a strongly evolving intrinsic equation of state combined with energy transfer between the dark sectors can generate an effective expansion history whose apparent dark-energy behavior differs substantially from that of the underlying fluid.

A particularly notable consequence of the interaction is the large shift in the inferred present-day matter density, $\Omega_{m0}=0.156^{+0.050}_{-0.077}$, compared with $0.3093\pm0.0055$ in the noninteracting thawing model. By contrast, the Hubble constant remains essentially unchanged, $H_0=67.43\pm0.56\,{\rm km\,s^{-1}\,Mpc^{-1}}$. The low value of $\Omega_{m0}$ reflects the strong degeneracy among the present-day matter abundance, the interaction strength, and the dark-energy evolution. Because the data combination considered here does not include the full range of low-redshift structure-growth information, the viability of this region should be tested against additional large-scale-structure observables.

The statistical improvement produced by the interaction is substantial. The interacting thawing model gives $\Delta\chi^2_{\rm MAP}=-10.00$, compared with $-3.93$ in the noninteracting case. Thus, introducing $\beta$ improves the MAP $\chi^2$ by approximately $6.1$. Even after penalizing the additional parameter, the AIC changes from $\Delta{\rm AIC}=-1.93$ to $\Delta{\rm AIC}=-6.00$. According to the conventional AIC criterion, the interacting thawing model therefore receives substantially greater support than its noninteracting counterpart and strong support relative to $\Lambda$CDM. This result demonstrates that, although the general $Iw_0w_a$CDM model does not require a nonzero coupling, an interaction can become statistically important when the intrinsic dark-energy evolution is restricted to a particular trajectory.

%Through the relation in Eq.~\eqref{eq:thawing_model}, this shift corresponds to a larger magnitude of  $w_a$ and hence a stronger time evolution of  $w(z)$ than in the noninteracting thawing model, as illustrated in Fig.~\ref{fig5}. Despite this stronger evolution,  $w_\mathrm{de}^\mathrm{eff}$  remains  above  $-1$ throughout the evolution, with  no phantom-divide crossing.  The interaction also produces a substantial downward shift in $\Omega_{m0}$, from $0.3093\pm0.0055$ to $0.156^{+0.050}_{-0.077}$,  whereas $H_0$ remains close to the noninteracting value.
 
% The interacting thawing model substantially improves the best fit,  with $\Delta\chi^2_{\rm MAP}=-10.00$, approaching the improvement seen in the $w_0w_a$CDM  model. After accounting for the additional coupling parameter,   $\Delta{\rm AIC}$ decreases from $-1.93$ for the noninteracting thawing model to $-6.00$,  indicating increased statistical support for the interacting extension. The thawing trajectory therefore provides an example in which a dark-sector interaction can improve the best fit within a restricted dynamical dark-energy model.  

\begin{figure}[!htbp]
\centering
\includegraphics[width=0.8\columnwidth]{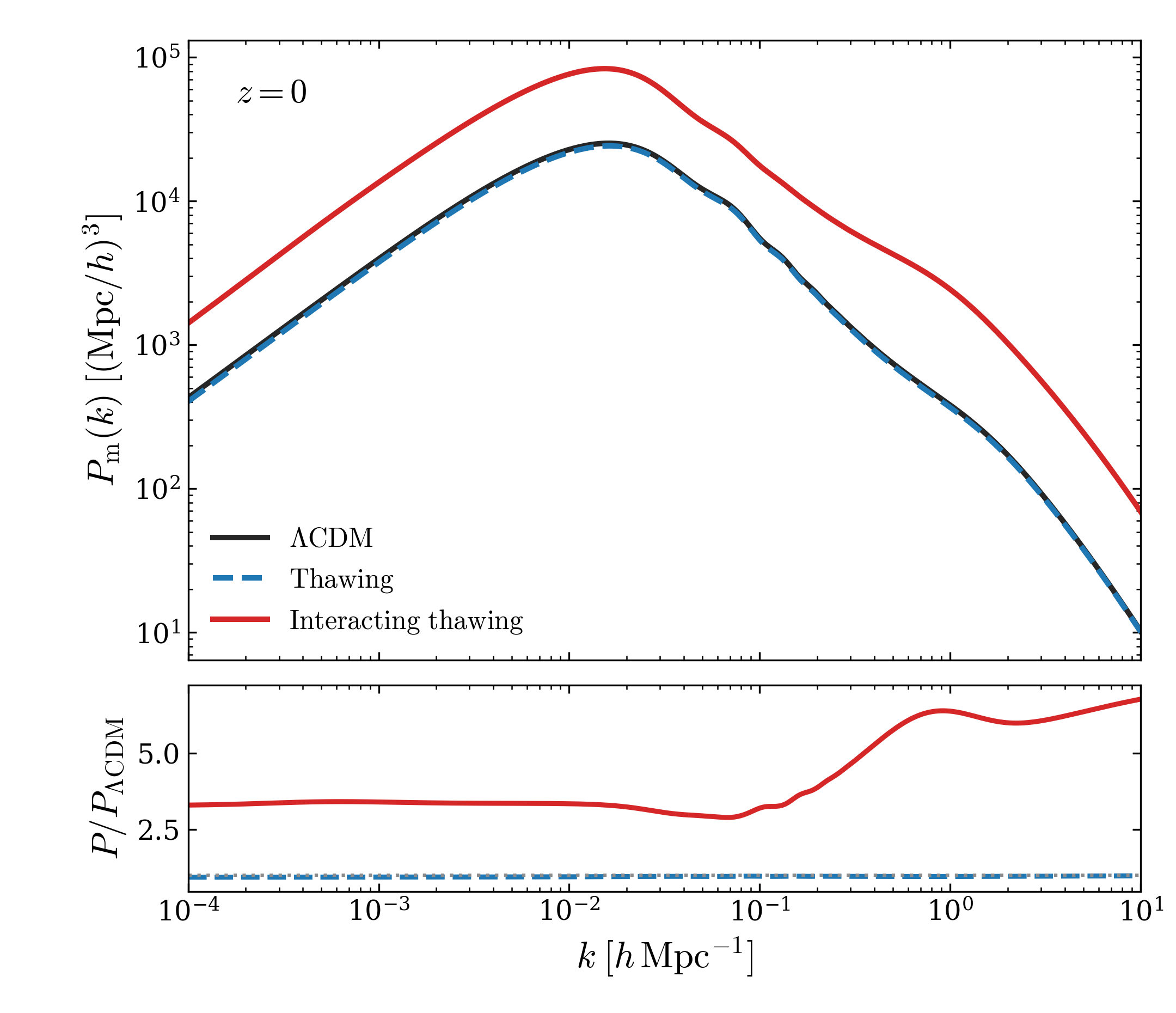}
\caption{Present-day matter power spectra for the thawing model and its interacting extension.}
\label{fig6}
\end{figure}

The interacting thawing solution also produces a distinctive signature in structure growth. Figure~\ref{fig6} shows the present-day matter power spectrum. The noninteracting thawing prediction remains close to that of $\Lambda$CDM, consistent with the relatively modest departure of its expansion history from the standard model. In contrast, the interacting thawing model exhibits a pronounced enhancement of matter power over the scales shown.

% Fig.~\ref{fig6} demonstrates that the interacting thawing model produces a much stronger modification of the matter power spectrum than that produced by the noninteracting thawing model. While the latter  remains close to the $\Lambda$CDM prediction, the interacting thawing model exhibits  a notable enhancement of power. This behavior  contrasts sharply with the suppression found in the $Iw_0w_a$CDM model and can be traced primarily to the opposite sign of the coupling parameter $\beta$. As indicated by Eq.~\eqref{eq:delta_c_ide}, a positive $\beta$ enhances the growth of the CDM density contrast on subhorizon scales, which in turn increases the total matter perturbation and hence the matter power spectrum $P_{m}(k)$.  

The sign of this effect is consistent with the positive coupling preferred by the interacting thawing fit. From Eq.~\eqref{eq:delta_c_ide}, the interaction-dependent contribution to the CDM perturbation equation is approximately $\beta{\cal H}\frac{\rho_{\rm de}}{\rho_c}\delta_{c,{\bf k}}$ on subhorizon scales. For $\beta>0$, this term acts in the direction of enhanced CDM perturbation growth. The interaction simultaneously modifies the background evolution of $\rho_c$, so the resulting change in $P_m(k)$ reflects both the altered CDM abundance and the modified growth of $\delta_c$. For the preferred interacting thawing parameters, their combined effect is a substantial enhancement of the present-day matter power spectrum.

This large growth enhancement is particularly important for assessing the model. The same interaction that improves the fit to the CMB+BAO+SNIa data also drives the cosmological parameters into a region characterized by a low present-day matter fraction and unusually strong matter clustering. The interacting thawing solution should therefore be regarded as a sharply testable prediction of the restricted model rather than as evidence that this parameter region is already favored by all cosmological observations. Incorporating low-redshift probes of structure growth, such as redshift-space distortions, galaxy clustering, and weak gravitational lensing, will be important for determining whether the improved fit to the data considered here survives once growth information is included.

As in the $Iw_0w_a$CDM case, the power spectrum displayed in Figure~\ref{fig6} corresponds to the representative parameter set adopted for the calculation. A more complete assessment would propagate the cosmological posterior into the predicted distribution of $P_m(k)$ or related growth observables. Nevertheless, the contrast between Figures~\ref{fig3} and \ref{fig6} already demonstrates an important feature of interacting dynamical-dark-energy models: the direction and magnitude of the modification to structure growth depend sensitively on both the dark-energy trajectory and the sign of the preferred interaction.

\subsection{Mirage dark energy}
\label{subsec:mirage_results}

We now examine the mirage trajectory, which describes evolving equations of state that can  closely mimic the distance-redshift relation of $\Lambda$CDM over an extended redshift range~\citep{Linder2007Mirage}. Following the DESI extended dark-energy analyses~\citep{Lodha2024PhysicsDE,DESI2025ExtendedDE}, we approximate this trajectory  in the $w_0$-$w_a$ plane as 
\begin{eqnarray}
 w_a=-3.66(1+w_0).
\label{eq:mirage_model}
\end{eqnarray}
As in the thawing case, the intrinsic dark-energy equation of state is therefore described by a single independent parameter $w_0$, while the interacting extension introduces  the additional  coupling parameter $\beta$. The posterior distributions are shown in Fig.~\ref{fig7}, and the marginalized constraints are summarized in Tab.~\ref{tab:mirage_constraints}.

\begin{center}
\centering
\includegraphics[width=0.42\columnwidth]{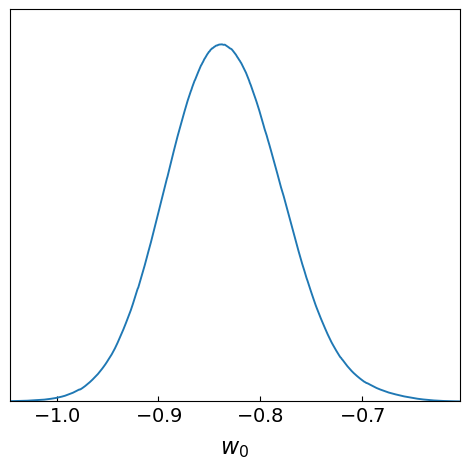}
\includegraphics[width=0.54\columnwidth]{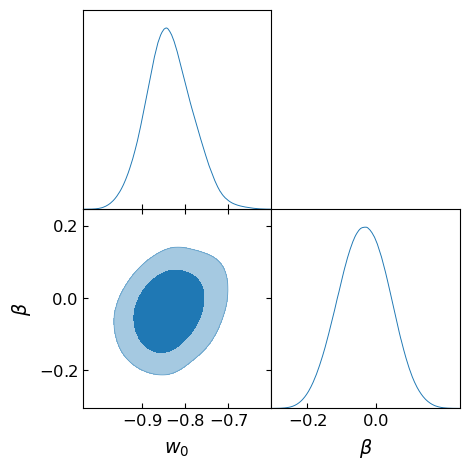}
\captionof{figure}{Posterior distributions of  selected parameters for the mirage model and its interacting extension.}
\label{fig7}
\end{center}

\begin{center}
\centering
\captionof{table}{Constraints  for the mirage model and its interacting extension.}
\label{tab:mirage_constraints}
\small
\setlength{\tabcolsep}{3.5pt}
\begin{tabular}{lcc}
\hline\hline
Parameter & Mirage & Interacting mirage \\
\hline
$w_0$ & $-0.835\pm0.056$ & $-0.837^{+0.050}_{-0.057}$ \\
$\beta$ & $-$ & $-0.034\pm0.074$ \\
$H_0$ & $67.67\pm0.37$ & $67.44\pm0.52$ \\
$\Omega_{m0}$ & $0.3099\pm0.0047$ & $0.320\pm0.020$ \\
%$S_8$ & $0.8211\pm0.0072$ & $0.815^{+0.017}_{-0.019}$ \\
\hline\hline
$\Delta\chi^2_{\rm MAP}$ & $-10.50$ & $-10.75$ \\
$\Delta{\rm AIC}$ & $-8.50$ & $-6.75$ \\
\hline\hline
\end{tabular}
\end{center}

\begin{figure}[!tbp]
\centering
\includegraphics[width=0.8\columnwidth]{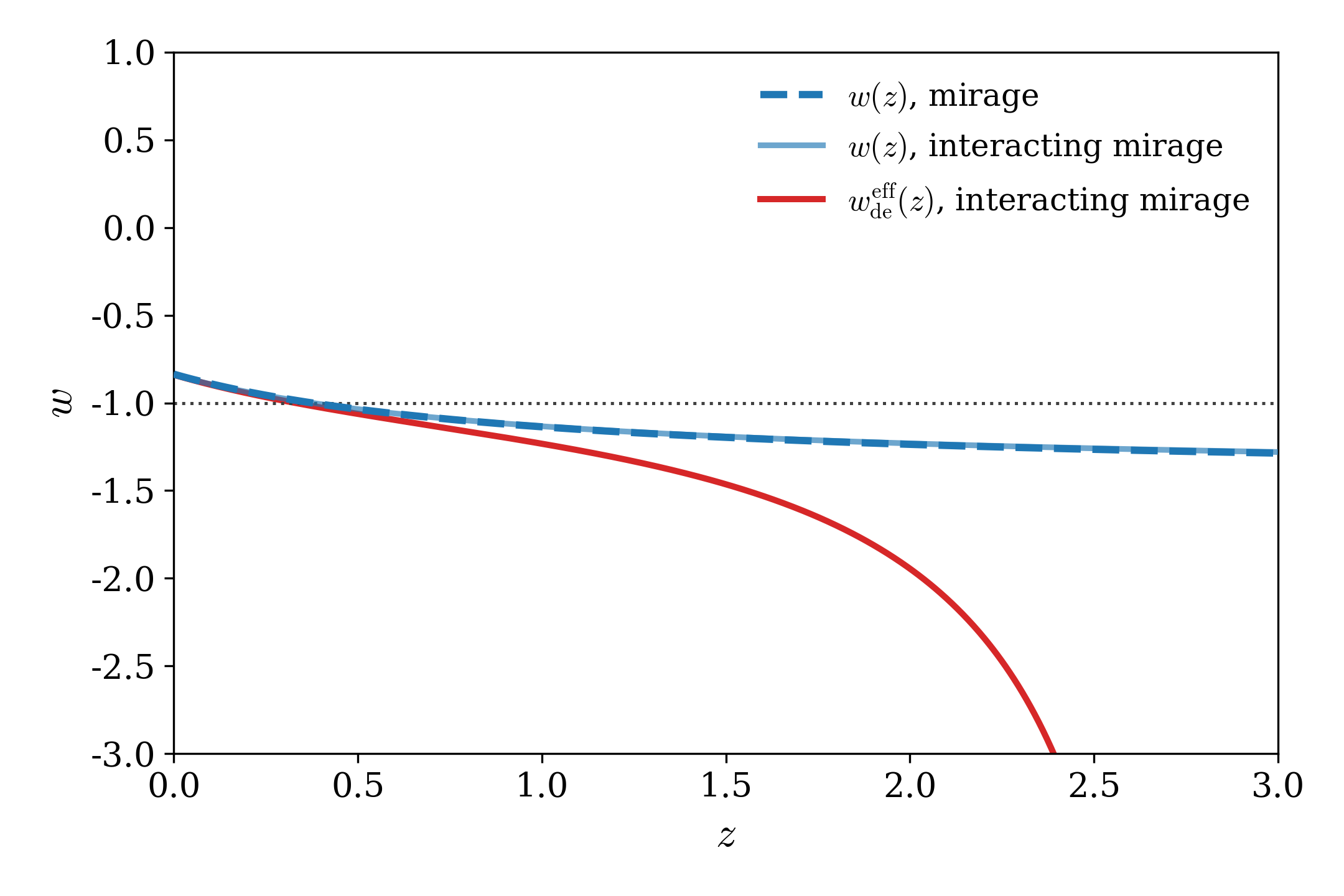}
\caption{Evolutions of the dark-energy equation of state for the mirage model and its interacting extension.}
\label{fig8}
\end{figure}

For the noninteracting mirage model,  we obtain $w_0=-0.835\pm0.056$. Through Eq.~\eqref{eq:mirage_model}, the posterior mean corresponds to $w_a\simeq-0.60$. This value aligns well with the allowed range for $w_a$ in the $w_0w_a$CDM model. The inferred matter density, $\Omega_{m0}=0.3099\pm0.0047$, and Hubble constant, $H_0=67.67\pm0.37\,{\rm km\,s^{-1}\,Mpc^{-1}}$, are likewise consistent with those obtained in $w_0w_a$CDM.
%The constraints on $\Omega_{m0}$ and $H_0$ are also consistent with those presented in Tab.~\ref{tab1}. 
The mirage trajectory therefore passes close to the region of the $w_0$--$w_a$ plane  preferred by the general CPL model, while requiring only one independent dark-energy parameter.  As illustrated in Fig.~\ref{fig8}, the preferred $w(z)$ evolves from $w<-1$ at higher redshift to $w>-1$ at the present epoch and crosses the phantom divide at late times. 
 Relative to $\Lambda$CDM, the mirage model gives  $\Delta\chi^2_{\rm MAP}=-10.50$ and $\Delta{\rm AIC}=-8.50$. Its best-fit likelihood is therefore nearly as good as that of $w_0w_a$CDM. Because the mirage model achieves this improvement with only one additional parameter relative to $\Lambda$CDM, its AIC is lower than that of the two-parameter $w_0w_a$CDM model. According to the conventional AIC criterion, the mirage trajectory therefore receives strong support relative to $\Lambda$CDM for the data combination considered here.

% while its fewer free parameters lead to a slightly smaller AIC and hence a strong statistical preference for the mirage model.

The situation changes very little when the dark-sector interaction is introduced. We find
 $\beta=-0.034\pm0.074$, which is fully consistent with zero.  %at  $1\sigma$. 
 The dark-energy parameter $w_0=-0.837^{+0.050}_{-0.057}$ is essentially unchanged  from its noninteracting value, and the inferred  $H_0$ is similarly stable. The matter density shows a modest upward shift from $0.3099\pm0.0047$ to $0.320\pm0.020$, accompanied by a larger uncertainty, but remains consistent with the noninteracting value. Thus, in contrast to the interacting thawing model, 
 the introduction of $\beta$ neither substantially changes the preferred dark-energy evolution nor drives the background parameters into a qualitatively different region.
% introducing the interaction in the mirage model produces no substantial shifts in the background parameters.

This stability can be understood from the location of the mirage trajectory in the $w_0$--$w_a$ plane. For the preferred $w_0\simeq-0.84$, Eq.~\eqref{eq:mirage_model} gives $w_a\simeq-0.60$, placing the model close to the region already selected by the unrestricted $w_0w_a$CDM fit. There is therefore little need for the interaction to compensate for a mismatch between the assumed dark-energy trajectory and the expansion history preferred by the data. This behavior contrasts with the thawing trajectory, for which introducing the interaction substantially shifts both the dark-energy parameters and the inferred matter abundance.

Fig.~\ref{fig8}  further illustrates this result. The intrinsic $w(z)$ histories of the interacting and noninteracting mirage models are nearly indistinguishable because their preferred values of $w_0$ are essentially identical. At low redshift, the reconstructed $w_{\rm de}^{\rm eff}$ also remains close to the intrinsic equation of state. Deviations become more noticeable toward higher redshift as the interaction-induced modification of the effective dark-energy density accumulates. For the slightly negative posterior mean of $\beta$, this difference reflects energy transfer from dark energy to CDM, although the coupling itself is statistically consistent with zero.

%shows that the evolutionary behavior of $w(z)$ in the interacting mirage model closely resembles that of the noninteracting case. This similarity occurs because the allowed values of $w_0$ in both the noninteracting and interacting mirage models are nearly identical. At $z<0.5$,  the evolution of $w_{\rm de}^{\rm eff}$  also closely tracks  $w(z)$, whereas deviations become more apparent at higher redshifts. This behavior arises from the interaction contribution to the effective dark-energy density and, for the central value $\beta<0$, becomes increasingly important toward higher redshifts.

The statistical comparison reinforces this interpretation. Introducing the interaction changes the best-fit statistic only from $-10.50$ to $-10.75$. Thus, the additional coupling parameter improves the MAP $\chi^2$ by only $0.25$.  This small gain is insufficient to compensate for the extra degree of freedom: $\Delta{\rm AIC}$ increases from $-8.50$ for the noninteracting mirage model to $-6.75$ for its interacting extension.   The data therefore provide no statistical motivation for introducing an interaction once the dark-energy evolution is restricted to the mirage trajectory.

%leading to an increase in $\Delta{\rm AIC}$ from $-8.50$ to $-6.75$. The mirage trajectory therefore accounts for the preferred dark-energy evolution without supporting an additional dark-sector interaction.

This result provides a particularly clear illustration of the degeneracy between dark-energy dynamics and dark-sector interactions. The general $Iw_0w_a$CDM analysis showed that a coupling becomes unnecessary once sufficient freedom is allowed in $w(z)$. The mirage model demonstrates that the same conclusion can hold even within a one-parameter dark-energy model, provided that its trajectory passes sufficiently close to the region of expansion histories favored by the data. Conversely, the interacting thawing results show that a coupling can become statistically useful when the imposed dark-energy trajectory alone does not reproduce that region as efficiently. The apparent evidence for an interaction is therefore sensitive not simply to the number of dark-energy parameters, but to the particular form of the allowed dark-energy evolution.

\begin{figure}[!tbp]
\centering
\includegraphics[width=0.8\columnwidth]{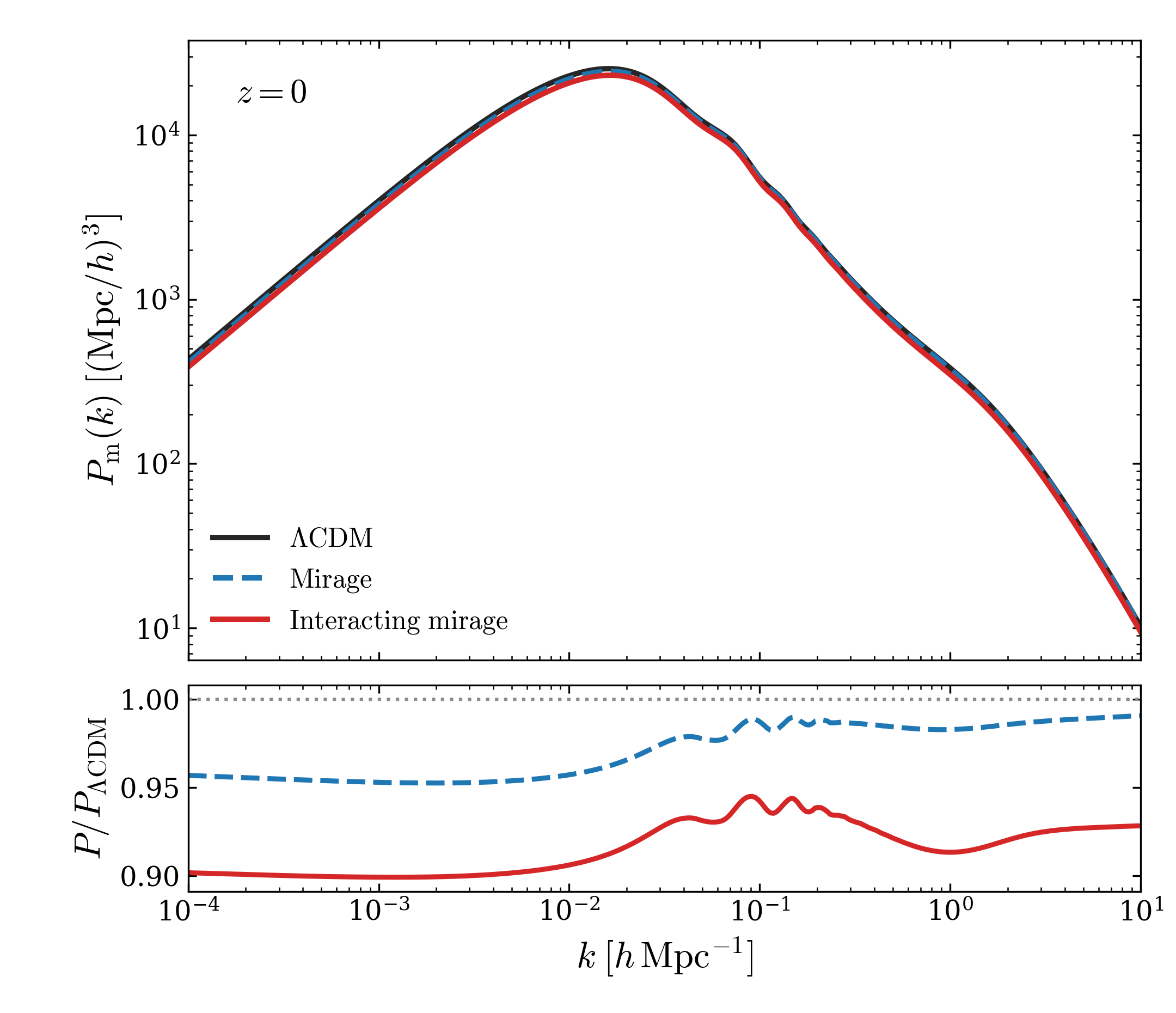}
\caption{Present-day matter power spectra for the mirage model and its interacting extension.}
\label{fig9}
\end{figure}

The structure-growth predictions provide a complementary comparison.  Figure~\ref{fig9} shows the present-day matter power spectrum for the mirage model and its interacting extension.  The noninteracting mirage model produces only a modest suppression relative to $\Lambda$CDM, remaining within approximately $5\%$ over the scales shown. Including the interaction increases the suppression to approximately $10\%$ for the representative parameter set used in the figure.

%demonstrates that the noninteracting mirage model suppresses the matter power spectrum by less than $5\%$ relative to $\Lambda$CDM over the scales considered. The suppression increases to approximately $10\%$ when the dark-sector interaction is included. Although the interaction therefore produces an additional modification of structure growth, the resulting difference may remain difficult to distinguish with current observations. 

For the slightly negative central value of $\beta$, the direction of this additional suppression is consistent with the behavior inferred from Eq.~\eqref{eq:delta_c_ide}: a negative interaction contribution tends to reduce the growth of the CDM density contrast on subhorizon scales. The magnitude of the effect is considerably smaller than in the $Iw_0w_a$CDM case, however, reflecting both the much smaller preferred coupling and the relatively modest changes in the background matter density.

Because $\beta=-0.034\pm0.074$ is consistent with zero, the additional suppression displayed by the representative interacting spectrum should not be interpreted as a statistically required departure from the noninteracting prediction. Rather, Figure~\ref{fig9} illustrates the growth modification associated with the parameter set used for the plotted curve. Propagating the full posterior into the predicted matter power spectrum would be required to determine the statistical significance of this difference.

The mirage results therefore provide the cleanest case among the restricted trajectories considered here in which the preferred dark-energy dynamics largely remove the need for an additional interaction. The model reproduces almost the full best-fit improvement of $w_0w_a$CDM with a single dark-energy parameter, while the coupling remains consistent with zero and yields only a marginal additional improvement in the likelihood. We next examine GEDE, for which the situation is qualitatively different: the noninteracting model remains close to the $\Lambda$CDM limit, whereas introducing the interaction substantially changes both the preferred dark-energy evolution and the inferred matter sector.

\subsection{GEDE}
\label{subsec:gede_results}

Finally, we consider the GEDE model, 
which describes a recent transition in the dark energy density rather than a linear trajectory in the $w_0$-$w_a$ plane~\citep{LiShafieloo2020PEDE,YangEtAl2021GEDE,Lodha2024PhysicsDE,DESI2025ExtendedDE}. The equation of state for GEDE can be expressed as
\begin{eqnarray}
 w(a)=-1-\frac{\delta}{3}
 \left[1-\tanh\left(\delta\ln\frac{a}{a_e}\right)\right],
\label{eq18}
\end{eqnarray}
where $\delta$ is a positive constant that controls the sharpness of the transition, and $a_e$ denotes the scale factor at which the dark energy density equals the total matter density. The GEDE reduces to the cosmological constant dark energy when $\delta=0$. In the interacting case, this equality point is solved self-consistently in the coupled background. The posterior distributions for model parameters are shown in Fig.~\ref{fig10}, and the marginalized constraints are summarized in Tab.~\ref{tab:gede_constraints}.

\begin{center}
\centering
\includegraphics[width=0.42\columnwidth]{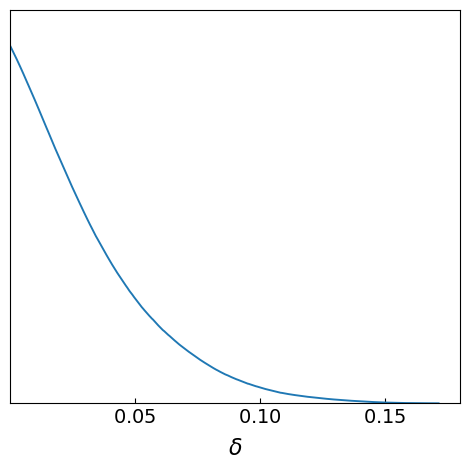}
\includegraphics[width=0.54\columnwidth]{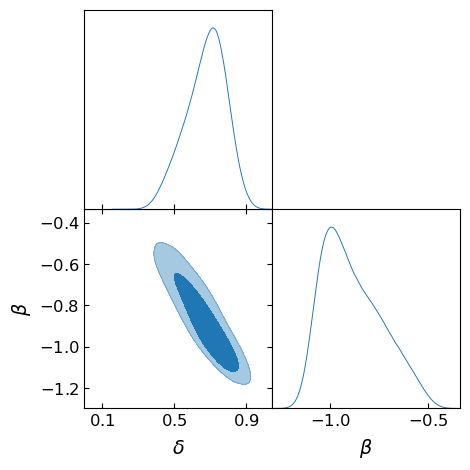}
\captionof{figure}{Posterior distributions for the GEDE model and its interacting extension.}
\label{fig10}
\end{center}

\begin{center}
\centering
\captionof{table}{Constraints for the GEDE model and its interacting extension.}
\label{tab:gede_constraints}
\small
\setlength{\tabcolsep}{3.5pt}
\begin{tabular}{lcc}
\hline\hline
Parameter & GEDE & Interacting GEDE \\
\hline
$\delta$ & $0.0309^{+0.0258}_{-0.0309}$ & $0.677^{+0.130}_{-0.093}$ \\
$\beta$ & $-$ & $-0.89^{+0.11}_{-0.19}$ \\
$H_0$ & $68.54^{+0.29}_{-0.34}$ & $67.78\pm0.52$ \\
$\Omega_{m0}$ & $0.3006\pm0.0036$ & $0.468^{+0.030}_{-0.012}$ \\
%$S_8$ & $0.8160\pm0.0071$ & $0.709^{+0.011}_{-0.015}$ \\
\hline\hline
$\Delta\chi^2_{\rm MAP}$ & $0.00$ & $-8.86$ \\
$\Delta{\rm AIC}$ & $+2.00$ & $-4.86$ \\
\hline\hline
\end{tabular}
\end{center}

For the GEDE model, we obtain $\delta=0.0309^{+0.0258}_{-0.0309}$. The posterior therefore extends to the $\Lambda$CDM limit, $\delta=0$, and provides no significant evidence for an emergent dark-energy evolution. Correspondingly, the best-fit likelihood is essentially identical to that of $\Lambda$CDM, $\Delta\chi^2_{\rm MAP}=0.00$. Because GEDE introduces one additional parameter, this gives $\Delta{\rm AIC}=+2.00$. The noninteracting GEDE model is therefore not favored over $\Lambda$CDM by the data combination considered here.

%Figure~\ref{fig10} illustrates that observational data provide only an upper limit on $\delta$, leading to an expansion history  close to that of  $\Lambda$CDM. This is also reflected in the nearly constant $w(z)=-1$ depicted in Fig.~\ref{fig11}. As a result, the GEDE model provides no improvement in the best fit, with $\Delta\chi^2_{\rm MAP}=0.00$, while $\Delta{\rm AIC}=+2.00$ slightly favors $\Lambda$CDM.

The inferred background parameters are $H_0=68.54^{+0.29}_{-0.34}\,{\rm km\,s^{-1}\,Mpc^{-1}}$ and $\Omega_{m0}=0.3006\pm0.0036$. The value of $H_0$ is somewhat higher than in the CPL, thawing, and mirage models considered above, but remains well below the local distance-ladder determination. Since $\delta$ is consistent with zero, the preferred noninteracting GEDE expansion history remains close to that of $\Lambda$CDM.

Introducing a dark-sector interaction changes the result dramatically. We obtain $\delta=0.677^{+0.130}_{-0.093}$ and $\beta=-0.89^{+0.11}_{-0.19}$.  The interacting model therefore favors both substantial dark-energy evolution and a large negative coupling. Under our sign convention, $\beta<0$ corresponds to energy transfer from dark energy to CDM. The interaction is accompanied by a large shift in the present-day matter density, $\Omega_{m0}=0.468^{+0.030}_{-0.012}$, while the Hubble constant decreases to $H_0=67.78\pm0.52\,{\rm km\,s^{-1}\,Mpc^{-1}}$. Thus, as in the interacting thawing case, allowing energy exchange within the dark sector opens a region of parameter space with a matter abundance very different from that inferred in the corresponding noninteracting model.

%Allowing for the interaction shifts the transition parameter to $\delta=0.677^{+0.130}_{-0.093}$ and yields $\beta=-0.89^{+0.11}_{-0.19}$, indicating a strong preference for a negative coupling and hence energy transfer from dark energy to CDM. The interaction also shifts $\Omega_{m0}$ upward,  from $0.3006\pm0.0036$ to $0.468^{+0.030}_{-0.012}$, while $H_0$ decreases moderately from $68.54^{+0.29}_{-0.34}$ to $67.78\pm0.52\,{\rm km\,s^{-1}\,Mpc^{-1}}$. The interacting GEDE model yields $\Delta\chi^2_{\rm MAP}=-8.86$ and $\Delta{\rm AIC}=-4.86$, showing that the improvement in the fit more than compensates for the additional coupling parameter. Thus, unlike the noninteracting GEDE model, the interacting extension receives moderate statistical support relative to $\Lambda$CDM.

\begin{figure}[!tbp]
\centering
\includegraphics[width=0.8\columnwidth]{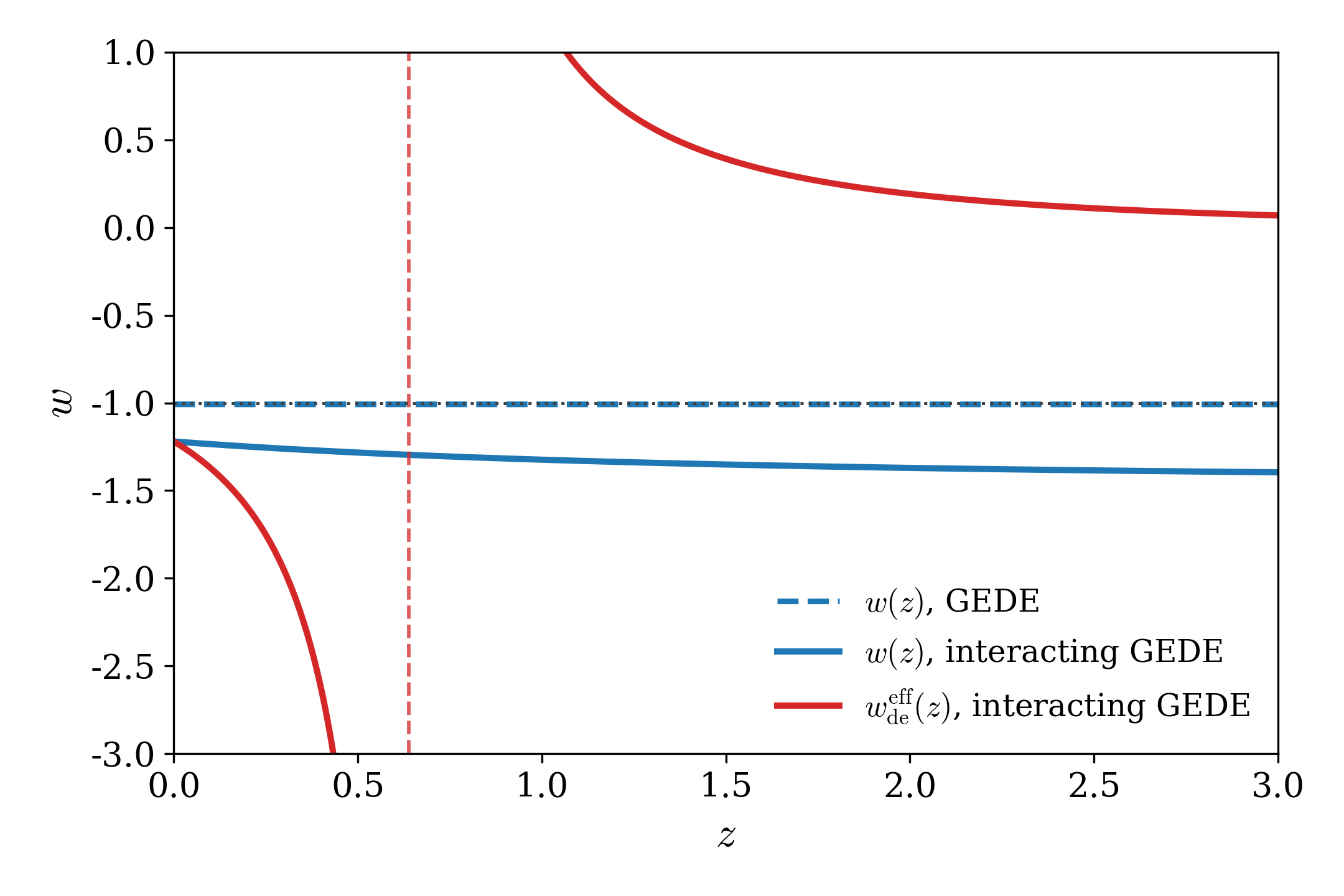}
\caption{Evolutions of the intrinsic dark-energy equation of state  for the GEDE model and its interacting extension. The vertical dashed line marks $\rho_{\rm de}^{\rm eff}=0$.}
\label{fig11}
\end{figure}

 Figure~\ref{fig11} shows the evolution of the intrinsic and effective dark-energy equations of state.
 For positive $\delta$, Eq.~\eqref{eq18} gives $w<-1$, with the departure from $-1$ becoming more pronounced toward earlier times. The large value of $\delta$ preferred in the interacting model therefore corresponds to substantially stronger intrinsic dark-energy evolution than in the noninteracting GEDE case. At the same time, the negative coupling modifies the evolution of both $\rho_c$ and $\rho_{\rm de}$, so that the reconstructed $w_{\rm de}^{\rm eff}$ can differ markedly from the intrinsic $w(z)$. This again illustrates that similar background observables need not imply a unique decomposition into dark matter and dark energy.
  
%As shown in Fig.~\ref{fig11}, the large value of $\delta$ in the interacting model produces a clearly phantom equation of state over the plotted redshift range. For the preferred negative coupling, energy transfer from dark energy to CDM reduces the reconstructed $\rho_{\rm de}^{\rm eff}$ at earlier times. At the redshift marked by the vertical dashed line, $\rho_{\rm de}^{\rm eff}$ crosses zero, causing $w_{\rm de}^{\rm eff}$ to develop a pole and change sign across it. This pole is a feature of the effective noninteracting reconstruction and does not represent a physical singularity of the underlying interacting model.

The interaction also leads to a substantial improvement in the fit. For interacting GEDE, we find
 $\Delta\chi^2_{\rm MAP}=-8.86,$ compared with $\Delta\chi^2_{\rm MAP}=0.00$ in the noninteracting model. Thus, introducing the single interaction parameter improves the MAP $\chi^2$ by $8.86$. After accounting for the two additional parameters relative to $\Lambda$CDM, the interacting model gives $\Delta{\rm AIC}=-4.86.$  The interaction therefore changes GEDE from a model that receives no support over $\Lambda$CDM according to AIC to one that is moderately favored relative to $\Lambda$CDM for the data combination considered here.

This result should be contrasted with those obtained for the other dark-energy trajectories. In the mirage model, the noninteracting trajectory already closely reproduces the expansion history preferred by the data, and adding $\beta$ improves the MAP $\chi^2$ by only $0.25$. In the thawing model, the interaction improves it by approximately $6.1$. For GEDE, the corresponding improvement is $8.86$. The statistical role of the interaction therefore depends strongly on the assumed intrinsic dark-energy evolution: it is unnecessary for a trajectory already well aligned with the preferred expansion history, but can become important when the noninteracting trajectory is more restrictive.
 
\begin{figure}[!tbp]
\centering
\includegraphics[width=0.8\columnwidth]{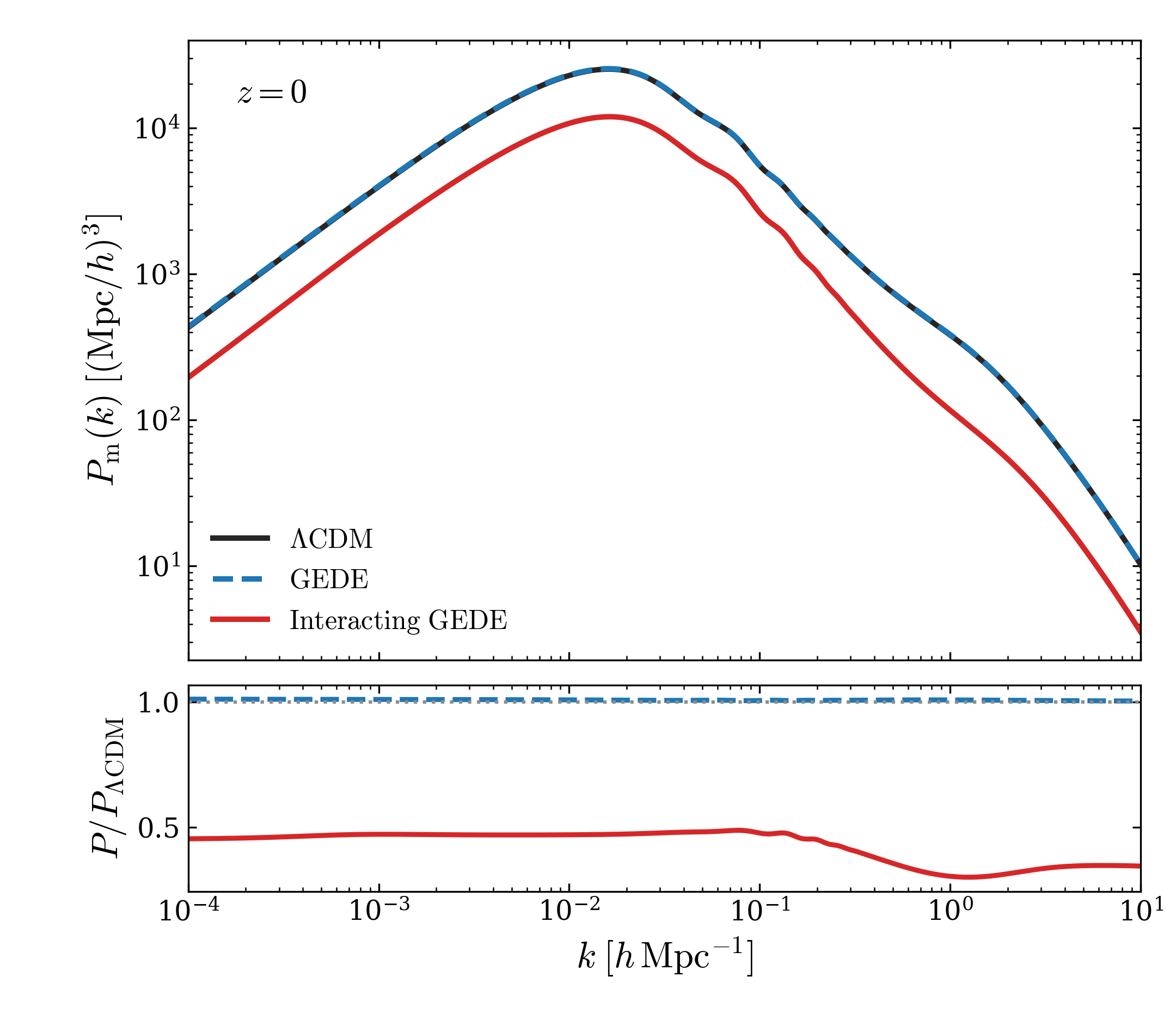}
\caption{Present-day matter power spectra for the GEDE model and its interacting extension.}
\label{fig12}
\end{figure}

The interacting GEDE solution, however, has particularly striking consequences for structure formation. 
Figure~\ref{fig12}  shows the present-day matter power spectrum.  The noninteracting GEDE  prediction remains close to that of $\Lambda$CDM, as expected from its small preferred value of $\delta$.  In contrast, the interacting model produces a very strong suppression of matter power, with the amplitude reduced by roughly a factor of two relative to $\Lambda$CDM over much of the range shown.

The direction of this effect is consistent with the large negative coupling. As discussed in Sec.~\ref{subsec:results_iw0wa}, on subhorizon scales the interaction-dependent contribution to the CDM perturbation equation is approximately $\beta{\cal H}\frac{\rho_{\rm de}}{\rho_c}$. For $\beta<0$, this term suppresses the growth of the CDM density contrast. In the interacting GEDE model, the magnitude of the preferred coupling is much larger than in either $Iw_0w_a$CDM or interacting mirage, so the corresponding modification of the perturbation evolution is also much stronger. Together with the substantial change in the background CDM history, this produces the pronounced suppression of $P_m(k)$ shown in Figure~\ref{fig12}.

This growth prediction is an important qualification to the improvement in the CMB+BAO+SNIa fit. The large present-day matter fraction, $\Omega_{m0}\simeq0.47$, and the strong suppression of matter clustering represent substantial departures from the standard cosmological solution. Although the data combination analyzed here permits this region and yields an improved likelihood, the model must also be consistent with direct low-redshift measurements of structure formation. Galaxy clustering, redshift-space distortions, weak gravitational lensing, and cluster-abundance measurements should therefore provide powerful tests of the interacting GEDE solution. In particular, the factor-of-two-level modification of the representative matter power spectrum suggests that structure-growth data could strongly constrain the parameter region favored by the background-dominated data combination used here.

As in the previous subsections, the spectrum in Figure~\ref{fig12} should be interpreted according to the parameter set used to generate the curve. A posterior-predictive treatment of $P_m(k)$ or observables such as $f\sigma_8$ and weak-lensing statistics would be required to quantify the range of growth histories allowed by the full posterior. This is especially important for interacting GEDE because the growth modification is sufficiently large that structure data may substantially alter the inferred posterior.

The GEDE results complete a clear pattern across the dynamical-dark-energy models considered in this work. The interaction is not generically preferred once dark-energy dynamics are allowed. Instead, its inferred role depends strongly on the trajectory imposed on $w(z)$. The mirage trajectory already follows the region favored by the general CPL fit and requires essentially no interaction. The thawing and GEDE trajectories, by contrast, acquire substantially improved fits when an interaction is introduced, but they favor couplings of opposite signs and consequently predict qualitatively different structure-growth histories: enhanced clustering for interacting thawing and strongly suppressed clustering for interacting GEDE. These contrasting predictions provide a direct means of breaking the background-level degeneracy between dark-energy dynamics and dark-sector interactions.

% power spectrum nearly coincides with that of $\Lambda$CDM. However, the interaction significantly suppresses  power, reducing $P_m(k)$ by approximately a factor of two over  a wide range of scales. This substantial effect is primarily attributed to the preferred large negative coupling, which inhibits the growth of the CDM density contrast, as described by Eq.~\eqref{eq:delta_c_ide}. Thus, although the interacting GEDE model provides an improved fit to the data considered here, it predicts a pronounced modification of structure growth that could provide an important test with large-scale-structure observations.

To summarize the statistical role of the interaction across the different dark-energy trajectories, Table~\ref{tab:interaction_summary} compares the improvement obtained when the coupling parameter $\beta$ is added to each corresponding noninteracting model.

\begin{table}[!tbp]
\centering
\caption{Statistical impact of introducing the dark-sector interaction for the different dynamical-dark-energy models. Here $\Delta\chi^2_{\rm int}$ denotes the change in the minimum $\chi^2$ relative to the corresponding noninteracting model, rather than relative to $\Lambda$CDM.}
\label{tab:interaction_summary}
\begin{tabular}{lcc}
\hline\hline
Dark-energy model & $\Delta\chi^2_{\rm int}$ & Coupling constraint \\
\hline
$w_0w_a$ & $-0.03$ & $-0.35^{+0.77}_{-0.62}$ \\
Mirage & $-0.25$ & $-0.034\pm0.074$ \\
Thawing & $-6.07$ & $0.53^{+0.24}_{-0.13}$ \\
GEDE & $-8.86$ & $-0.89^{+0.11}_{-0.19}$ \\
\hline\hline
\end{tabular}
\end{table}

The contrast is clear. Introducing an interaction produces essentially no improvement for $w_0w_a$CDM or the mirage trajectory, whereas it substantially improves the fit for the thawing and GEDE trajectories. Moreover, the latter two models favor couplings of opposite signs. This comparison demonstrates that the inferred statistical role of the interaction depends strongly on the assumed intrinsic dark-energy evolution.

\section{Conclusions}
\label{sec:conclusion}

In this work, we have reassessed the evidence for interactions between CDM and dynamical dark energy in light of DESI DR2. We considered an interaction of the form $Q=\beta H\rho_{\rm de}$ and confronted the resulting models with P-ACT CMB temperature, polarization, and lensing data, DESI DR2 BAO measurements, and DES-Dovekie SNIa. By employing the generalized PPF framework, we consistently evolved cosmological perturbations across the phantom divide and were therefore able to explore interacting dynamical-dark-energy models without restricting the equation of state to remain entirely in either the phantom or nonphantom regime.

We first considered the general $Iw_0w_a$CDM model, allowing both CPL dark-energy dynamics and a dark-sector interaction. We find that the coupling is consistent with zero, $\beta=-0.35^{+0.77}_{-0.62}$, while the preference for an evolving dark-energy equation of state persists. Most notably, adding the interaction improves the MAP $\chi^2$ by only $0.03$ relative to the noninteracting $w_0w_a$CDM model. The AIC correspondingly changes from $\Delta{\rm AIC}=-6.72$ for $w_0w_a$CDM to $-4.75$ for $Iw_0w_a$CDM. Thus, once sufficient freedom is allowed in the intrinsic dark-energy evolution, the present data provide no statistical motivation for an additional interaction. This result is consistent with the possibility that part of the apparent coupling preference found in more restricted interacting models reflects a degeneracy between energy exchange in the dark sector and intrinsic dark-energy dynamics.

To investigate how this conclusion depends on the assumed form of $w(z)$, we then considered three one-parameter dynamical-dark-energy trajectories: thawing, mirage, and GEDE. The results differ markedly among them. The mirage trajectory passes close to the region of the $w_0$--$w_a$ plane preferred by the general CPL fit and reproduces nearly the same improvement in likelihood with only one dark-energy parameter. Introducing an interaction then yields only a marginal additional improvement, $\Delta\chi^2_{\rm MAP}\simeq-0.25$, with $\beta=-0.034\pm0.074$ fully consistent with zero. The mirage model therefore provides a particularly clear example in which an appropriately aligned dark-energy trajectory leaves little statistical role for an additional dark-sector interaction.

The thawing and GEDE models exhibit qualitatively different behavior. For the thawing trajectory, introducing the interaction improves the MAP $\chi^2$ by approximately $6.1$ and shifts the posterior toward a positive coupling, $\beta=0.53^{+0.24}_{-0.13}$, corresponding to energy transfer from CDM to dark energy in our convention. For GEDE, the noninteracting model remains close to the $\Lambda$CDM limit and provides essentially no improvement in the best fit, whereas its interacting extension improves the MAP $\chi^2$ by $8.86$ and favors a large negative coupling, $\beta=-0.89^{+0.11}_{-0.19}$, corresponding to energy transfer from dark energy to CDM. The interaction can therefore become statistically important when the assumed noninteracting dark-energy trajectory is more restrictive, but both the magnitude and direction of the preferred energy transfer depend strongly on that trajectory.

These results demonstrate that evidence for a dark-sector interaction cannot be assessed independently of assumptions about dark-energy dynamics. More specifically, our results suggest that when the allowed $w(z)$ trajectory already follows the region of expansion histories preferred by the data, as in $w_0w_a$CDM and the mirage model, introducing an interaction yields little additional improvement. For more restrictive trajectories, such as thawing and GEDE, the interaction can compensate for differences in the background evolution and substantially improve the fit. The resulting coupling preference should therefore be interpreted jointly with the assumed dark-energy parameterization rather than as model-independent evidence for energy exchange in the dark sector.

The background degeneracy is accompanied by sharply different predictions for structure formation. The representative $w_0w_a$CDM and mirage models remain relatively close to $\Lambda$CDM in their present-day matter power spectra, whereas their interacting extensions produce additional suppression. More strikingly, the positive coupling favored in the interacting thawing model substantially enhances matter clustering, while the large negative coupling favored in interacting GEDE strongly suppresses it. The corresponding shifts in the inferred matter abundance are also substantial, with $\Omega_{m0}\simeq0.16$ for interacting thawing and $\Omega_{m0}\simeq0.47$ for interacting GEDE. Thus, models that provide comparably good fits to CMB, BAO, and SNIa data can predict radically different late-time matter distributions.

These extreme growth predictions also provide an important qualification to the statistical improvements obtained for the interacting thawing and GEDE models. The present analysis does not incorporate the full range of low-redshift large-scale-structure information, and an improved fit to the data combination considered here does not by itself establish the viability of these parameter regions. Measurements of redshift-space distortions, galaxy clustering, weak gravitational lensing, and cluster abundance can directly test the predicted growth histories and may substantially tighten the interaction constraints. Future analyses combining DESI full-shape clustering with weak-lensing surveys such as DES, HSC, KiDS, Rubin, Euclid, and Roman should therefore provide substantially greater power to distinguish intrinsic dark-energy evolution from interactions within the dark sector.

Our conclusions are necessarily conditional on the phenomenological framework considered here. We have adopted the specific interaction $Q=\beta H\rho_{\rm de}$ with the energy-momentum transfer parallel to the CDM four-velocity, together with several particular parameterizations of the intrinsic dark-energy evolution. Other interaction forms or momentum-transfer prescriptions can lead to different background and perturbation dynamics. It will therefore be important to determine whether the trajectory dependence identified here persists across a broader class of interacting models.

Overall, our analysis shows that the apparent evidence for dark-sector interactions in the DESI DR2 era is strongly intertwined with assumptions about the dynamics of dark energy. Allowing sufficiently flexible or suitably aligned dark-energy evolution can largely remove the statistical preference for an interaction, whereas more restrictive trajectories can favor substantial couplings of either sign. At the same time, these background-level degeneracies give rise to markedly different predictions for structure growth. Joint analyses of expansion-history and large-scale-structure observables will therefore be essential for determining whether the evolving dark sector indicated by current data is better described by intrinsic dark-energy dynamics, energy exchange between dark matter and dark energy, or a combination of the two.

\paragraph{Note added.}
While this work was being completed, Ref.~\citep{Yang2026BeyondDDE} appeared, presenting an independent analysis of interacting dynamical-dark-energy models with DESI DR2. Although the interaction prescription and model assumptions differ from those adopted here, that study likewise finds that allowing additional freedom in dark-energy dynamics can weaken the inferred evidence for a dark-sector interaction.

\begin{acknowledgments}
This work was supported in part by the NSFC under Grant  Nos. 12275080 and 12635002,  and the Innovative Research Group of Hunan Province under Grant No.~2024JJ1006.

\end{acknowledgments}

\bibliographystyle{aasjournalv7}
\bibliography{CF}

\end{document}